\documentclass[twocolumn,aps,superscriptaddress,nofootinbib]{revtex4-2}
\usepackage{graphicx}
\usepackage{amsmath}
\usepackage{amssymb}
\usepackage{slashed}
\usepackage{xcolor}
\usepackage[normalem]{ulem}

\usepackage[colorlinks]{hyperref}  

\begin{document}
\title{Cabibbo suppressed hyperon production off nuclei induced by antineutrinos}
\author{M. Benitez Galan}
\email{mbenitezgalan@ugr.es}
\affiliation{Departamento de F\'isica
At\'omica, Molecular y Nuclear, 
Facultad de Ciencias, 
Universidad de Granada, E-18071, Granada,
Spain}
\author{L. Alvarez-Ruso}
\affiliation{Instituto de F\'isica Corpuscular
  (IFIC), Consejo Superior de Investigaciones
  Cient\'ificas (CSIC), E-46980 Paterna, Valencia,
Spain}
\author{M. Rafi Alam}
\affiliation{Department of Physics,
Aligarh Muslim University, Aligarh-202 002,
India}
\author{I. Ruiz Simo}
\email{ruizsig@ugr.es}
\affiliation{Departamento de F\'isica
At\'omica, Molecular y Nuclear and Instituto
Interuniversitario Carlos I de F\'isica
Te\'orica y Computacional, 
Facultad de Ciencias, 
Universidad de Granada, E-18071, Granada,
Spain}
\author{M.J. Vicente Vacas}
\affiliation{Departamento de F\'isica
  Te\'orica and Instituto de F\'isica
  Corpuscular (IFIC), Centro Mixto UVEG-CSIC,
  Valencia E-46071, Spain}

\begin{abstract}
In this work we study the production of $\Sigma$ and $\Lambda$
hyperons in strangeness changing $\Delta S = -1$ charged current
interactions of muon antineutrinos on nuclear targets. At the nucleon
level, besides quasielastic scattering we consider the inelastic
mechanism in which a pion is produced alongside the hyperon. Its
relevance for antineutrinos with energies below 2 GeV is conveyed in
integrated and differential cross sections. We observe that the
distributions on the angle between the hyperon and the final lepton
are clearly different for quasielastic and inelastic
processes. Hyperon final state interactions, modeled with an
intranuclear cascade, lead to a significant transfer from primary
produced $\Sigma$'s into final $\Lambda$'s. They also cause
considerable energy loss, which is apparent in hyperon energy
distributions. We have investigated $\Lambda$ production off
${}^{40}$Ar in the conditions of the recently reported MicroBooNE
measurement.  We find that the $\Lambda \pi$ contribution, dominated
by $\Sigma^*(1385)$ excitation, accounts for about one third of the
cross section.
\end{abstract}

\maketitle

\section{Introduction}\label{sect:intro}
The flavour changing charged currents of the Standard Model connect up
and strange quarks, $W^- \, u\rightarrow s$. Weak interactions on
nucleons can then lead to strange hadrons and, in particular, hyperons
in the final state: $W^- \, N\rightarrow Y$. The amplitudes for such
$\Delta S = -1$ processes are proportional to the sine of the Cabibbo
angle~\cite{Cabibbo:1963yz} or $V_{us}=0.22500 \pm
0.00067$~\cite{ParticleDataGroup:2022pth} and are therefore suppressed
with respect to their counterparts involving only $u \leftrightarrow
d$ transitions with $V_{ud}=0.97435 \pm
0.00016$~\cite{ParticleDataGroup:2022pth}. Nevertheless, the form
factors that characterize the electroweak $N\rightarrow Y$ current
encode relevant details about the hadronic structure, SU(3) breaking
corrections, G-parity or time-reversal invariance violation. They are
also a valuable input for the determination of CKM matrix elements.

On the theoretical side there has been a continuing effort trying to
obtain the relevant axial and vector form
factors~\cite{Cabibbo:1965zza,LlewellynSmith:1971uhs,Cabibbo:2003cu}
and the possible SU(3) breaking effects from QCD and QCD-inspired
models. For instance, lattice
QCD~\cite{Shanahan:2015dka,Sasaki:2012ne,Sasaki:2017jue}, chiral
perturbation
theory~\cite{Zhu:2000zf,Lacour:2007wm,Ledwig:2014rfa,Shanahan:2015dka,Sauerwein:2021jxb},
$1/N_c$
expansions~\cite{Flores-Mendieta:1998tfv,Buchmann:2002et,CalleCordon:2012xz}
or quark
models~\cite{Schlumpf:1994fb,Ramalho:2015jem,Yang:2015era,Liu:2022ekr}
have been used. Nonetheless, our knowledge of these form factors is
still not satisfactory, in part due to the scarcity of experimental
information, which comes mostly from hyperon semileptonic decays and
is thus restricted to very small momentum transfers.

On the other hand, the reaction $ \bar{\nu}+N\rightarrow l^+ + Y$,
which better probes the momentum transfer dependence of the
$N\rightarrow Y$ form factors has been insufficiently
explored. Indeed, at the moment, there is just a handful of $\Lambda$
and $\Sigma$ production events observed at several bubble chambers:
Gargamelle, filled with freon~\cite{Eichten:1972bb,Erriquez:1977tr}
and propane~\cite{Erriquez:1978pg}, ANL with
deuterium~\cite{Barish:1974ye}, BNL with
hydrogen~\cite{Fanourakis:1980si,Baker:1981tx}, Fermilab with a heavy
neon-hydrogen mixture~\cite{Ammosov:1986jn} or with
deuterium~\cite{Son:1983xh} and SKAT, filled with
freon~\cite{SKAT:1989nel}. In spite of the low statistics and incoming
flux uncertainties, these experiments could estimate low-energy
($E_\nu < 20$~GeV) cross sections for $\Delta S =-1$ single $\Lambda$,
$\Sigma$, $\Delta S =0$ $\Lambda K$ production and $Y X$, $Y K X$,
where $X$ stands for additional hadrons. At higher energies, some
cross sections and rates, but also hyperon yields and polarization
measurements have been performed with bubble chambers (see
Ref.~\cite{Formaggio:2012cpf} for a comprehensive list of references)
and by the NOMAD experiment~\cite{NOMAD:2000wdf,NOMAD:2004djf}.

Recently, the first measurement of $\bar\nu_\mu + \, ^{40}\mathrm{Ar}
\rightarrow \mu^+ + \Lambda + X$, where $X$ denotes the final state
content without strangeness, has been reported by the MicroBooNE
Collaboration~\cite{MicroBooNE:2022bpw}. So far only five $\Lambda$
events have been identified analyzing the exposure of the MicroBooNE
liquid argon detector to the off-axis NUMI beam at FNAL. Fortunately,
the situation is bound to improve with the large data sample collected
by MicroBooNE and still awaiting analysis~\cite{MicroBooNE:2022bpw}
and the much larger one expected at the SBND
detector~\cite{Brailsford:2017rxe,Machado:2019oxb}.

The fact that the constituents of the baryonic matter present in
(anti)neutrino detectors are light $u$ and $d$ quarks, while the
corresponding antiquarks are only present in the sea of $q \bar{q}$
pairs, implies that $\Delta S = -1$ $W^- \, u\rightarrow s$ processes
induced by $\bar\nu$ are very different from the $\Delta S = 1$, $W^+
\, \bar u\rightarrow \bar s$ ones induced by $\nu$. In particular,
only antineutrino interactions can trigger the production of
single-hyperons and strange baryon resonances. The consequence of this
$\nu / \bar\nu$ asymmetry for CP-violation searches has not been
investigated although it has been pointed out that, owing to their
weak decays, $Y \rightarrow \pi N$, hyperons become an additional
source of pions in $\bar\nu$
interactions~\cite{Singh:2006xp,RafiAlam:2013prd,Fatima:2021ctt}. Furthermore,
if sufficiently determined by experiments, hyperon production could be
used to constrain the $\bar\nu$ contamination in a $\nu$
beam~\cite{MicroBooNE:2022bpw}.

As apparent from the experimental outline above, $\bar\nu$ scattering
on single nucleons leads to a variety of final states. First, starting
from low energies, there are $\Delta S = -1$ quasielastic (QE)
interactions, $ \bar{\nu} + N\rightarrow l^+ + Y$. The formalism to
write the QE cross section in terms of transition form factors,
relating the later to the nucleon ones by flavor SU(3) rotations, was
laid down in Ref.~\cite{Cabibbo:1965zza} and followed closely in
subsequent
studies~\cite{Singh:2006xp,Mintz:2006yp,Kuzmin:2008zz}. Second, with a
slightly higher energy threshold, an additional pion could be
produced, $ \bar{\nu} + N\rightarrow l^+ + Y + \pi + X$. For this
process, which we label as $Y\pi$, a model in terms of light
pseudoscalar mesons and baryon octets as effective degrees of freedom
was proposed in Ref.~\cite{Dewan:1981ab}. It has been improved and
extended to include the lowest-lying decuplet resonances in
Ref.~\cite{BenitezGalan:2021jdm}. This study finds that the
$\Sigma(1385)\,3/2^+$ intermediate state plays a prominent role in
$\Lambda \pi$ production. A specific reaction mechanism in which the
$\Lambda(1405)\,1/2^-$ is excited, contributing to the $\Sigma \pi$
final state has also been considered~\cite{Ren:2015bsa}. In addition,
two and more pions can accompany the hyperon as the energy further
increases. Another possible $\Delta S = -1$ process is $ \bar{\nu} + N
\rightarrow l^+ + N' + \bar{K}$~\cite{Dewan:1981ab,Alam:2011vwg}; it
can lead to hyperon production in nuclei (introduced below) through
$\bar K \, N \rightarrow Y \pi$ final state interactions. Next we
have, $\Delta S = 0$, $ \bar{\nu} + N \rightarrow l^+ + Y + K$. It has
been theoretically studied in Born
approximations~\cite{Shrock:1975an,Adera:2010zz} and, more recently,
accounting for unitarization in coupled
channels~\cite{Nakamura:2015rta}. Let us recall that the production of
hyperons by $\Delta S=-1$ channels (QE, $Y\pi$) is suppressed as
compared to the $\Delta S=0$ ones by a factor $(V_{us}/V_{ud})^2
\approx 0.05$. However, this reduction is more than compensated at low
energies, $E_{\bar\nu} < 2$~GeV, by the fact that the associated
production of hyperons requires the creation of an additional kaon and
is, therefore, strongly reduced by phase space.

Neutrino experiments rely on heavy targets to increase the available
statistics. Indeed, most hyperons produced in the laboratory arise
from scattering on nuclear targets, which introduces significant
complications for the analysis and interpretation of experimental
results. In this instance, the weak interaction takes place on a bound
nucleon from a nuclear target while the scattering process leaves the
residual nucleus in an excited state. The initial nucleon has been
described with the global~\cite{Thorpe:2020tym} and
local~\cite{Singh:2006xp,Sobczyk:2019uej,Thorpe:2020tym} Fermi gas
approximations and taking the nuclear mean field and nucleon-nucleon
correlations into account~\cite{Sobczyk:2019uej}. Additionally, it
must be considered that hyperons produced in the primary scattering
propagate through the nucleus undergoing final state interactions
(FSI), where they can collide with the nucleons, change direction,
lose energy or even turn into a different hyperon species before
getting out of the nucleus and being detected.  Hyperons can also be
created by $\bar K N$ FSI, as mentioned above, or by high-energy pions
via $\pi \, N \rightarrow Y \, K$ secondary scattering.  These FSI,
which can be handled with semiclassical
methods~\cite{Singh:2006xp,Sobczyk:2019uej,Thorpe:2020tym,Lalakulich:2012gm},
have strong impact on the observables, sizably exceeding the influence
of the nuclear initial state treatment~\cite{Sobczyk:2019uej}.

Unlike QE hyperon production, $Y\pi$ mechanisms have not yet been
studied in nuclei. In this context, it is worth noticing that there is
no QE $\Sigma^+$ production on a single nucleon. This hyperon could
only appear due to nuclear FSI.  However, allowing for the presence of
additional pions, namely the $Y\pi$ channel, the $\Sigma^+$ hyperon
can be directly produced on a single proton.

In the present study of the $\Lambda$ and $\Sigma$ production off
nuclei induced by charged current interactions of muon antineutrinos,
the inelastic $Y \pi$ channel is considered alongside the QE one. By
restricting ourselves to the $E_{\bar{\nu}} \lesssim 2$ GeV range of
Laboratory energies, which is probed at MicroBooNE and SBND, we can
neglect hyperon production accompanied by more than one pion, the
associated $Y K$ reaction channel and secondary hyperon production
induced by $\bar K$. We include FSI accounting for the propagation of
hyperons in the nuclear medium with the help of a Monte Carlo
simulation. The model is applied to the recent MicroBooNE
measurement~\cite{MicroBooNE:2022bpw}. We find that under the phase
space restrictions imposed by detection thresholds, the $Y \pi$
mechanism becomes particularly important.

The structure of this article is as follows: in
Sec.~\ref{sec:formalism} we present the formalism employed to describe
QE and $Y\pi$ primary reactions on nucleons and nuclei; hyperon FSI
are discussed in Sec.~\ref{sec:fsi}. Various observables: cross
sections, energy and angular distributions are described in
Sec.~\ref{sec:results}, together with our analysis of the MicroBooNE
measurement. Finally, in Sect.~\ref{sec:conclusions} we summarize the
main conclusions of this study.

\section{Formalism}\label{sec:formalism}

In this section, we describe the formalism employed to obtain the
total and differential cross sections for QE and inelastic ($Y \pi$)
hyperon production on nucleons and nuclei\footnote{Most of the
formalism is readily available
elsewhere~\cite{Singh:2006xp,BenitezGalan:2021jdm}, where more details
can be found. Here, for the sake of clarity, we provide a brief
outline unifying the notation of previous works.}. As nuclear effects,
our approach includes Fermi motion of the target nucleons and FSI of
the emitted hyperons.  Fermi motion is accounted for using the Fermi
gas model with the local density approximation.  FSI is implemented
within the model of intranuclear re-interaction of the primary
produced hyperons with the nucleons of the nuclei described in
Ref.~\cite{Singh:2006xp}~\footnote{A very similar approach was adopted
in Ref.~\cite{Sobczyk:2019uej}.}.

\subsection{Quasielastic hyperon production on nucleons}
\label{subsec:qe}

This reaction is the main source of strange baryons at low energies
with antineutrino beams.  As a consequence of the $\Delta S = \Delta
Q$ selection rule, the allowed channels are
\begin{eqnarray}
\bar{\nu}_l(k) + p(p) &\rightarrow&
 l^{+}(k^\prime)+ \Lambda(p_Y)\,, \label{eq:QEreactions1}\\
 \bar{\nu}_l(k) + p(p) &\rightarrow&
 l^{+}(k^\prime)+ \Sigma^0(p_Y)\,,
 \label{eq:QEreactions2}\\
 \bar{\nu}_l(k) + n(p) &\rightarrow&
 l^{+}(k^\prime)+ \Sigma^-(p_Y)\,,
 \label{eq:QEreactions3}
\end{eqnarray}
where $k, p, k^\prime, p_Y$ denote the antineutrino, nucleon, lepton
and hyperon four-momenta, respectively.

The cross section on a free nucleon can be written as
\begin{eqnarray}
d\sigma^{\mathrm{QE}}_N &=& \frac{1}{(2\pi)^2} \frac{1}{2 (s- m_N^2)}
 \delta^4(k+p-k^\prime-p_Y)\nonumber\\
 &&\frac{d^3k^\prime}{2
 E_l(\mathbf{k}^\prime)}\;
 \frac{d^3p_Y}{2E_Y(\mathbf{p}_Y)} 
\overline{\sum} \left|\mathcal{M}\right|^2,
 \label{eq: dxsectQEprimary}
\end{eqnarray}
where $s=(k+p)^2$ and $m_N$ stands for the nucleon mass. The outgoing
particle energies obey the on-shell conditions $E_Y = \sqrt{m_Y^2+
  \mathbf{p}_Y^2}$ and $E_l = \sqrt{m_l^2+ \mathbf{k^\prime}^2}$, in
terms of the corresponding masses $m_{Y,l}$ and three-momenta
$\mathbf{p}_Y$, $\mathbf{k^\prime}$.  The overlined summation symbol
denotes the sum over the final fermion polarizations and the average
over the initial ones.  The scattering amplitude matrix element is
given by
\begin{equation}\label{eq:Mel}
\mathcal{M}=\frac{G_F}{\sqrt{2}} \bar{v}(k)\gamma_\mu (1-\gamma^5)v(k^\prime) 
J_H^\mu \, 
 \end{equation}
where $\bar{v}$ and $v$ are the lepton spinors;
$G_F=1.1664\times10^{-5}$ GeV$^{-2}$~\cite{ParticleDataGroup:2022pth}
is the Fermi constant. The hadronic current
\begin{equation}
\label{eq:Jhad}
 J_H^\mu  = V_{us} \langle Y\left| \bar{s} \gamma^\mu (1-\gamma^5) u 
\right|N\rangle,
\end{equation}
can be expressed in terms of three vector and three axial-vector
transition form factors. Assuming SU(3) flavor symmetry, they can be
related to their nucleon counterparts. We follow the specific choices
of Ref.~\cite{Singh:2006xp} where the reader is referred for details.

\subsection{Hyperon-pion production on nucleons}\label{subsec:ypi}
The process under consideration now is generically given by
\begin{eqnarray}
\bar{\nu}_l(k) + N(p) \rightarrow
 l^{+}(k^\prime)+ \pi(p_m) +Y(p_Y), 
 \label{ypireaction}
\end{eqnarray}
where, in addition to the previously defined ones, $p_m$ denotes the
pion four-momentum. For $N=p$ the allowed $Y\pi$ final states are
$\Lambda\pi^0$, $\Sigma^0\pi^0$, $\Sigma^-\pi^+$ and $\Sigma^+\pi^-$
while for $N=n$, $Y\pi =$ $\Lambda\pi^-$, $\Sigma^0\pi^-$ and
$\Sigma^-\pi^0$.

The differential cross section for this reaction on free nucleons is
\begin{eqnarray}
d\sigma^{Y\pi}_N&=& \frac{1}{(2\pi)^5}\frac{1}{2 (s - m_N^2)}\delta^4(k+p-k'-p_Y-p_m)
 \nonumber\\
 &&\frac{d^3k^\prime}{2
 E_l(\mathbf{k}^\prime)}\;
 \frac{d^3p_m}{2E_m(\mathbf{p}_m)}\;
 \frac{d^3p_Y}{2E_Y(\mathbf{p}_Y)} 
 \overline{\sum}
 \left|\mathcal{M}\right|^2,\nonumber\\
 \label{eq: dxsectypiprimary}
\end{eqnarray}
where the matrix element is also given by Eq. \ref{eq:Mel} but with a
hadronic inelastic current
\begin{equation}
\label{eq:Jhad2}
 J_H^\mu  = V_{us} \langle Y \pi\left| \bar{s} \gamma^\mu (1-\gamma^5) u \, 
\right|N\rangle,
\end{equation}
which encompasses Born terms derived using the lowest order effective
Lagrangian including the low-lying baryon octet. The $N-Y$ vertices
incorporate the same transition form factors that characterize QE
hyperon production. The baryon decuplet enters via a $u$-channel
contribution of the $\Delta(1232)$ and the excitation of the
$\Sigma^*(1385)$ which decays to $\Sigma \pi$ and, predominantly, to
$\Lambda \pi$. The octet-to-decuplet form factors are expressed in
terms of the $N-\Delta(1232)$ ones using SU(3) symmetry. For full
details, the reader is referred to~\cite{BenitezGalan:2021jdm}.

\subsection{The nuclear cross section}

The cross section on a nucleus of mass number $A$ and charge $Z$ is
given by the integral over all possible nucleon momenta in the Fermi
sea
\begin{equation}
d\sigma^i_A = \int d^3r\, 2\int \frac{d^3p}{(2\pi)^3}\,
 n_N(\mathbf{p},\mathbf{r})\, d\sigma^i_N
\end{equation}
where
$n_N(\mathbf{p},\mathbf{r})=\theta(k^N_F(r)-\left|\mathbf{p}\right|)$
is the occupation number of the initial nucleon of type $N=p,n$ with
momentum $\mathbf{p}$. Its dependence on the radial coordinate comes
from the Fermi momentum
\begin{equation}
k^{N}_F(r)=\left[ 3\pi^2\rho_{N}(r) \right]^{1/3} \,,
\label{eq: dxsectQEnuclei}
\end{equation}
defined in terms of the local density of protons or neutrons. For the
proton densities we take empirical parametrizations from
Ref.~\cite{DeVries:1987atn}, re-scaling them by a factor of $(A-Z)/Z$
for neutrons.  Index $i$ stands for one of the possible QE or $Y\pi$
reactions. In the QE case only one of the nucleon types contributes
for a given hyperon species. In other words, $N=p,\,p,\,n$ for primary
produced $\Lambda,\,\Sigma^0\,,\Sigma^-$ respectively. In the case of
the $Y\pi$ reaction, for a given hyperon both protons and neutrons
contribute (with the charge of the outgoing pion changed accordingly)
except $\Sigma^+$ hyperons that are primary produced only on protons.

\subsection{Hyperon nuclear potentials} 
\label{sec:pot}

Hyperons are produced in the mean field of the nucleus. Such hyperon
nucleus potentials, and hyperon strong interactions in general, are
not very well known. There is a consensus in the literature that the
$\Lambda$ nuclear potential is attractive, with a depth of around
30~MeV~\cite{Vidana:1998ed,Rodriguez-Sanchez:2018oqv}. The $\Sigma$
potential is difficult to determine. Even its sign is not established
although repulsive values inside the nucleus are currently
preferred~\cite{Hirtz:2019rqe,Harada:2023otu}. We disregard the latter
but take into account the effect of a $\Lambda$ potential,
parametrized as
\begin{equation}
    \label{eq:Lpot}
    V_\Lambda (r) = - 30 \,\mathrm{MeV} \, \frac{\rho(r)}{\rho(0)} \,,
\end{equation}
with $\rho=\rho_p + \rho_n$, on the hyperon propagation through the
nucleus and FSI.

\subsection{Final State Interaction}\label{sec:fsi}

In a semiclassical language, hyperons, initially produced in QE or
$Y\pi$ processes on target nucleons, propagate through the nucleus
changing hyperon species and/or energy and direction via
$Y+N\rightarrow Y^\prime +N^\prime$ interactions. Next, we describe
the Monte Carlo simulation accounting for these FSI and the $\Lambda$
potential.

The propagation of a given hyperon, produced in one of the possible
primary interactions of a $\bar\nu$ with laboratory energy
$E_{\bar\nu}$, starts at a random position $\mathbf{r_0}$ inside the
nucleus. The momentum of the initial nucleon is generated
isotropically, with $|\mathbf{p}| < k^{N}_F(r)$. Those of the outgoing
lepton, $\mathbf{k'}$, and pion, $\mathbf{p_m}$, (if applicable) are
also randomly generated after energy conservation is imposed. The
hyperon momentum at this initial coordinate is constrained by momentum
conservation as $\mathbf{p_Y} = \mathbf{k} - \mathbf{k'} + \mathbf{p}$
(QE) or $\mathbf{p_Y} = \mathbf{k} - \mathbf{k'} + \mathbf{p} -
\mathbf{p_m}$ ($Y\pi$).  In the case of the $\Lambda$, this
$\mathbf{p_\Lambda}$ is regarded as the asymptotic momentum the
hyperon would have in the absence of FSI. To account for the potential
the $\Lambda$ initial energy is increased by $-V_\Lambda(r_0)$ and the
absolute value of its momentum is adjusted as
\begin{equation}
    \label{eq:pLambda}
    \mathbf{p_\Lambda}^2(\mathbf{r_0}) =  \left[ \sqrt{m_\Lambda^2+
  \mathbf{p}_\Lambda^2} - V_\Lambda(r_0) \right]^2 - m_\Lambda^2
\end{equation}
In this way, the $\Lambda$ hyperon is propagated as an on-shell
particle~\cite{Thorpe:2020tym}.  Functions
\begin{equation}
\frac{d \sigma_A^i}{d^3r d^3p d^3k^\prime} \,\, \left(
\frac{d \sigma_A^i}{d^3r d^3p d^3k^\prime d^3p_m} \right)
\nonumber    
\end{equation}
provide the weights assigned to the QE ($Y\pi$) events of channel $i$.

Once the initial properties of the event have been fixed, the
simulation proceeds by moving the hyperon along a trajectory
determined by the classical Hamilton equations of motion. For
$\Sigma$, the absence of potential implies that this trajectory is
just a straight line along the hyperon's momentum direction.  The
travelled distance $dl$ is chosen such that $P_Y dl\ll 1$. Here $P_Y$
is the probability per unit length of hyperon-nucleon interaction at
point $\mathbf{r}$, given by
\begin{equation}
    P_Y (r) = \sum_{f,f'} \left\{ \sigma_{Y n\rightarrow f}(\bar{s}) \rho_n(r) +
    \sigma_{Y p\rightarrow f'} (\bar{s})\rho_p(r) \right\} \,,
    \label{prob}
\end{equation}
where the sum is performed over all possible hyperon-nucleon final
states. The integrated elastic and quasielastic $YN$ cross sections
$\sigma_{Y N\rightarrow Y' N'}$ were extracted from the available
experimental data and conveniently parametrized in
Ref.~\cite{Singh:2006xp}.  They are evaluated at a $YN$ invariant
energy $\bar s$ averaged over the local Fermi sea. At this stage, a
random number $x\in [0,1]$ is generated. If $x > P_Y dl$ there is no
interaction and the hyperon moves further a distance $dl$.  If not, an
interaction has occurred and a new $Y'N'$ final state is randomly
chosen among the allowed ones according to their respective
probabilities. Next, the momentum of the new $Y'$ hyperon is
generated: from the $YN$ invariant mass, calculated for an initial
nucleon at rest, the center-of-mass (CM) momenta of the final hyperon
and nucleon are obtained; their (back to back) directions are randomly
generated assuming an isotropic CM angular distribution. Then, the two
momenta are boosted to the Laboratory frame. If the $N'$ momentum is
below the local Fermi level, the event is Pauli blocked: the
interaction does not actually take place and the original hyperon is
propagated a distance $dl$ through its trajectory. Otherwise, one has
a new hyperon $Y'$ with a new momentum.

Following Ref.~\cite{Thorpe:2020tym} we implement further adjustments
in this algorithm, which are required to account for the $\Lambda$
potential. In case of a $\Lambda N \rightarrow \Lambda N$ interaction
one should check that after the collision $\sqrt{m_\Lambda^2 +
  \mathbf{p}_\Lambda^2} + V_\Lambda(r) > m_\Lambda$. Otherwise, the
$\Lambda$ is trapped in the attractive potential, its propagation is
ceased and the hyperon is not counted as an asymptotic final
state.\footnote{A fraction of bound $\Lambda$ hyperons decays weakly
to $p \pi^-$ and could be experimentally measured from these decay
products. However, the distortion of the emitted pion and nucleon in
the nucleus will render the $\Lambda$ identification
difficult. Furthermore, detection thresholds will hinder the
measurement. In particular, as discussed below, low-energy $\Lambda$
are undetectable at MicroBooNE. Last but not least, one should recall
that a semiclassical cascade is unsuitable to describe the formation
and decay of $\Lambda$ hypernuclei.} Next, if a secondary $\Lambda$ is
born in a $\Sigma N \rightarrow \Lambda N'$ interaction, its energy is
increased by $-V_\Lambda(r)$ and its momentum re-adjusted according to
Eq.~\ref{eq:pLambda} (with $r$ instead of $r_0$), just as for a
primarily produced $\Lambda$ hyperon. Finally, after a $\Lambda N
\rightarrow \Sigma N'$ FSI, the $\Sigma$ energy has to be decreased by
$V_\Lambda(r)$ and its momentum adjusted to correspond to an on-shell
$\Sigma$ with the analogous of Eq.~\ref{eq:pLambda} for a $\Sigma$ at
position $\mathbf{r}$, unless $\sqrt{m_\Sigma^2 + \mathbf{p}_\Sigma^2}
+ V_\Lambda(r) < m_\Sigma$. In the latter case, no $\Sigma$ hyperon
can actually be created, so the interaction is disregarded and the
original $\Lambda$ continues its propagation.

Although the main purpose of this study is to evaluate the total
hyperon production, it would be of interest to consider as well the
FSI of pions produced via the $Y\pi$ processes. If such pions were
detected, it would lead to a better understanding of the underlying
physics.

The primary interaction occurs on the whole volume of the nucleus and,
thus, pions in their way out could scatter, changing energy,
direction, charge, or be absorbed. A comprehensive calculation of
these effects would require a cascade simulation for both hyperons and
pions. This is beyond of the scope of this work but the effect of pion
absorption can be reasonably estimated using an eikonal
approximation. This approach has been successfully used in the
analysis of several pion production processes in nuclei induced by
pions~\cite{Oset:1986qd} or
neutrinos~\cite{Singh:2006br,Zhang:2012xi}. In this approximation, the
probability for a pion to escape from the nucleus is given by
\begin{equation}
\label{eq:eikonal}
P_{\mathrm{no\, abs}}= \exp\left[-\frac{1}{\left|\mathbf{p_m} \right|}{\int^\infty_0
\mathrm{Im}\,\Pi_{\mathrm{abs}}\left(\mathbf{r}+\lambda\, \frac{\mathbf{p_m}}{\left|\mathbf{p_m} \right|}\right) \,d\lambda}
\right] 
\end{equation}
where $\mathbf{r}$ denotes the pion production point; ${\rm
  Im}\,\Pi_{\mathrm{abs}}$ is the fraction of the imaginary part of
the pion selfenergy in the nuclear medium, which corresponds to pion
absorption. For our estimates, we take this selfenergy from
Refs.~\cite{Oset:1987re,Salcedo:1987md}.

\section{Results}\label{sec:results}

We present several results for hyperon production off nuclei induced
by antineutrino beams. As discussed in the introduction, our
predictions for hyperon production are reliable as far as the $\Delta
S = -1$ QE and $Y\pi$ processes are prevalent. This restricts the
incoming neutrino energies to $E_{\bar\nu} \lesssim 2$~GeV. The
$\Delta S = 0$ is also suppressed for measurements where no kaons are
allowed in the final state.  Once the QE mechanism has been considered
in earlier studies, whose findings we support, here we focus on $Y\pi$
and the comparison between the two mechanisms.

For most of the results we have selected a light, $^{16}$O, and a
medium size, $^{40}$Ar, nucleus, both present in modern neutrino
detectors. Furthermore, we also discuss how the cross section depends
on the mass number. All results presented here are obtained for muonic
antineutrinos.

\subsection{Integrated cross sections}
\label{subsec:total_xsect}
\begin{figure*}[!ht]
\centering
\includegraphics[width=0.46\textwidth]{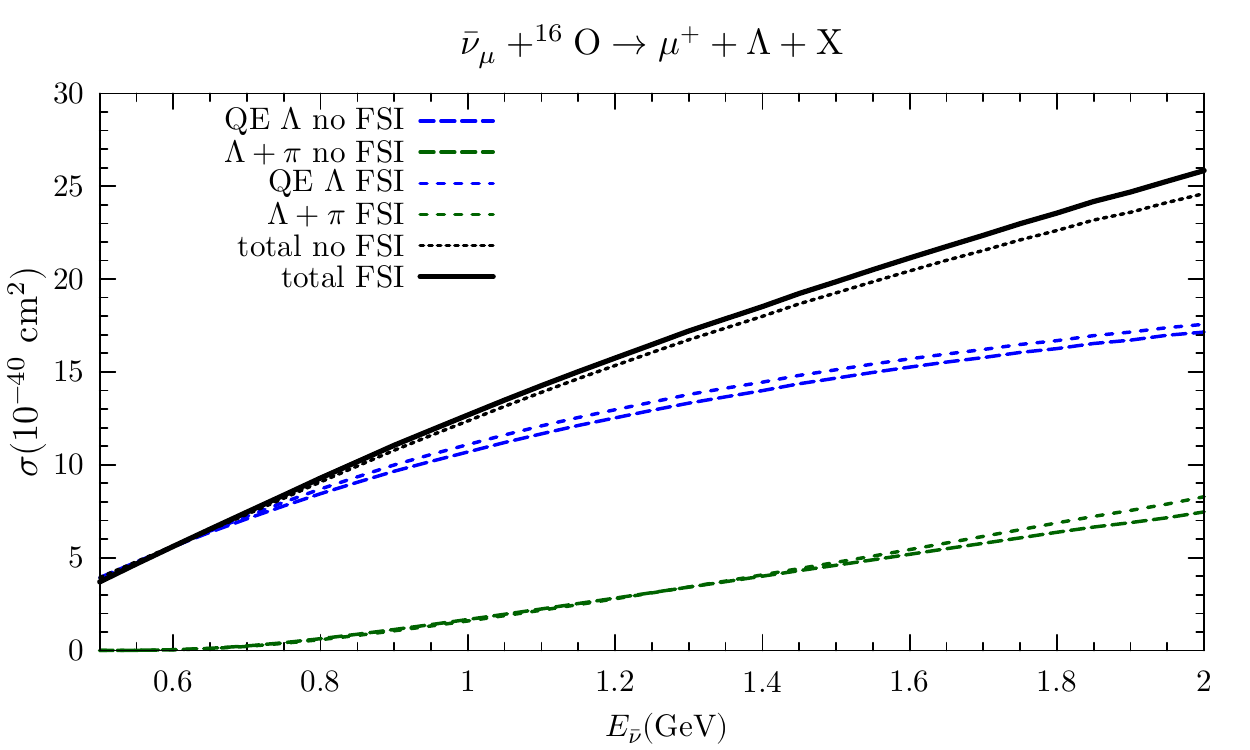}
\hspace{0.4cm}
\centering
\includegraphics[width=0.46\textwidth]{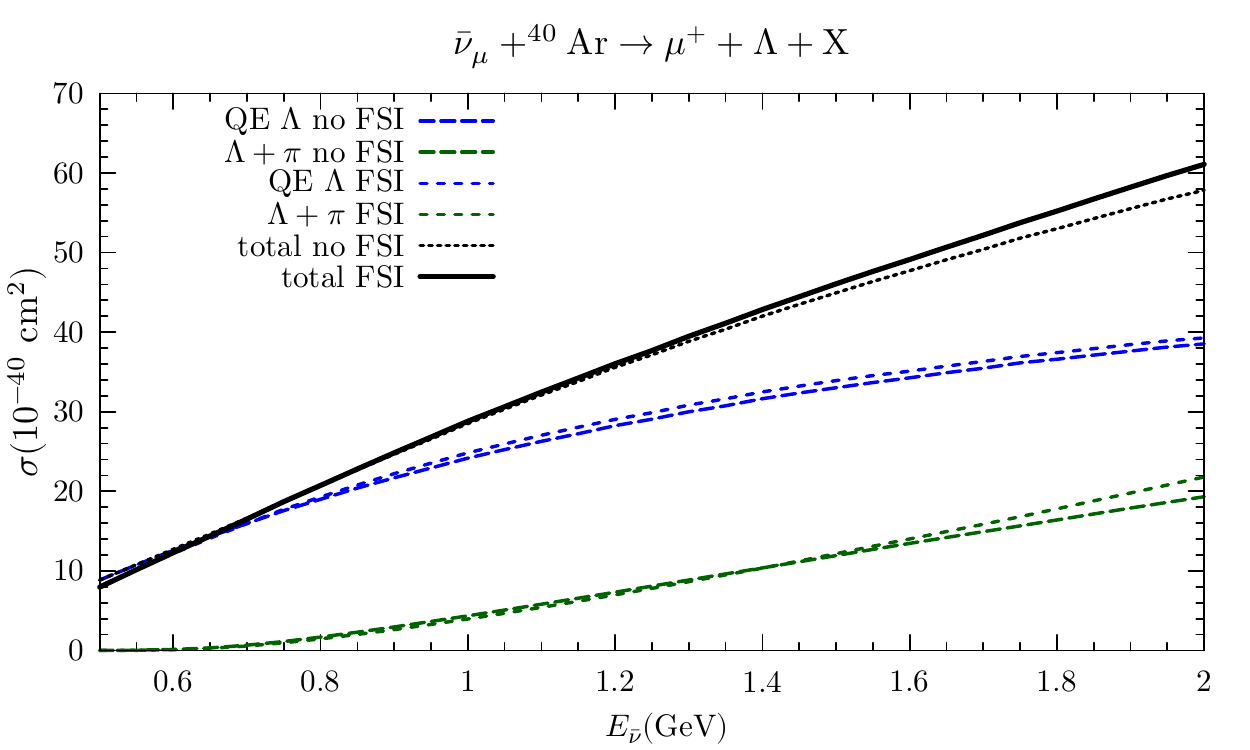}
\centering
\includegraphics[width=0.46\textwidth]{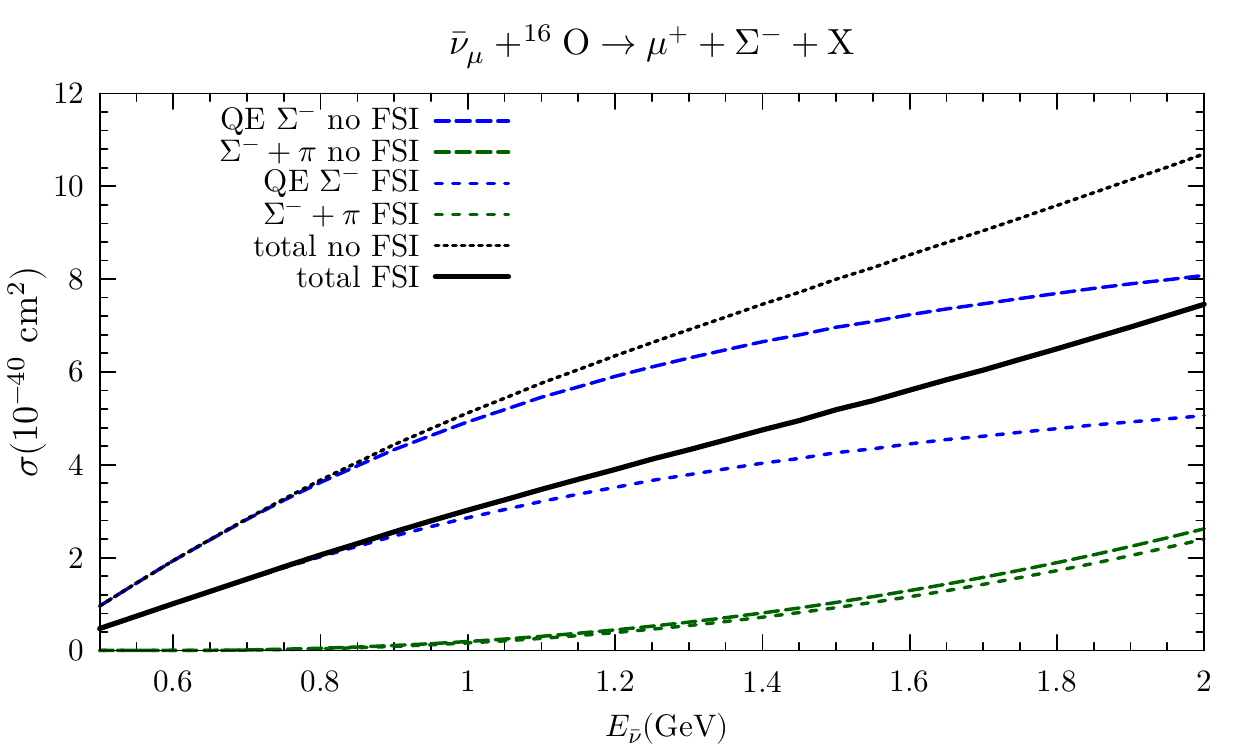}
\hspace{0.4cm}
\centering
\includegraphics[width=0.46\textwidth]{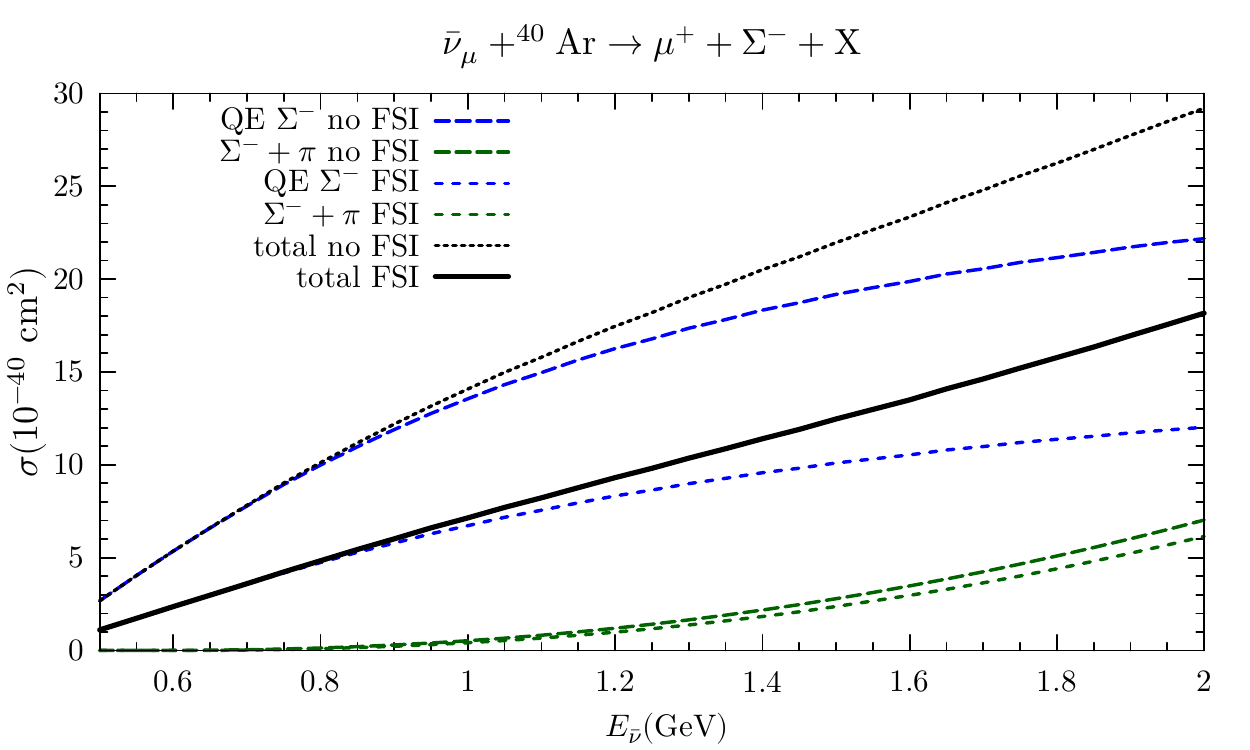}
\includegraphics[width=0.46\textwidth]{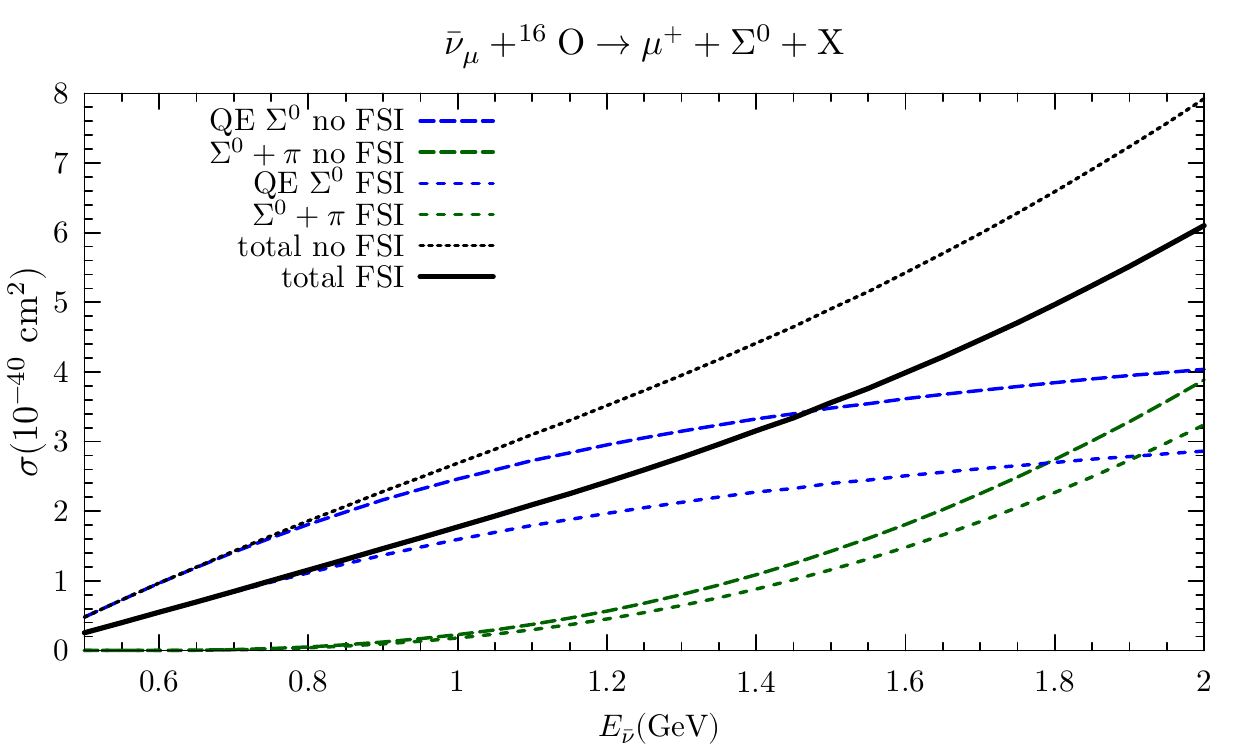}
\hspace{0.4cm}
\centering
\includegraphics[width=0.46\textwidth]{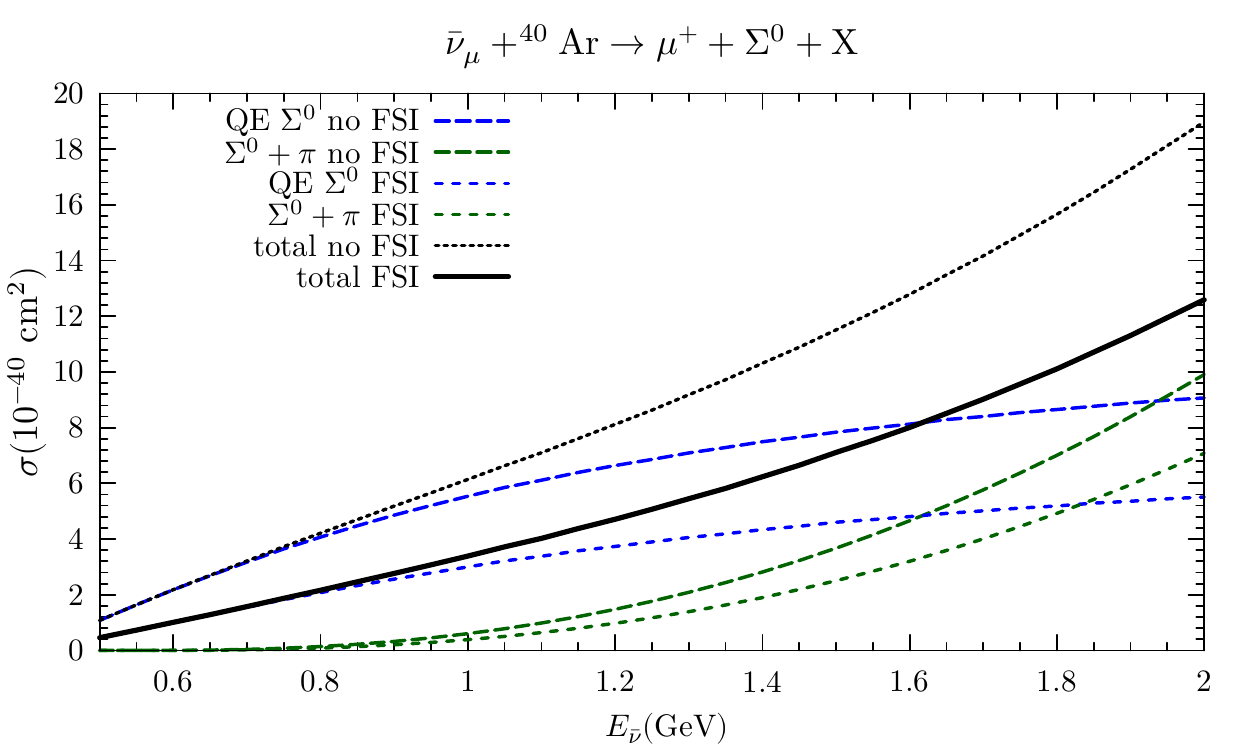}
\includegraphics[width=0.46\textwidth]{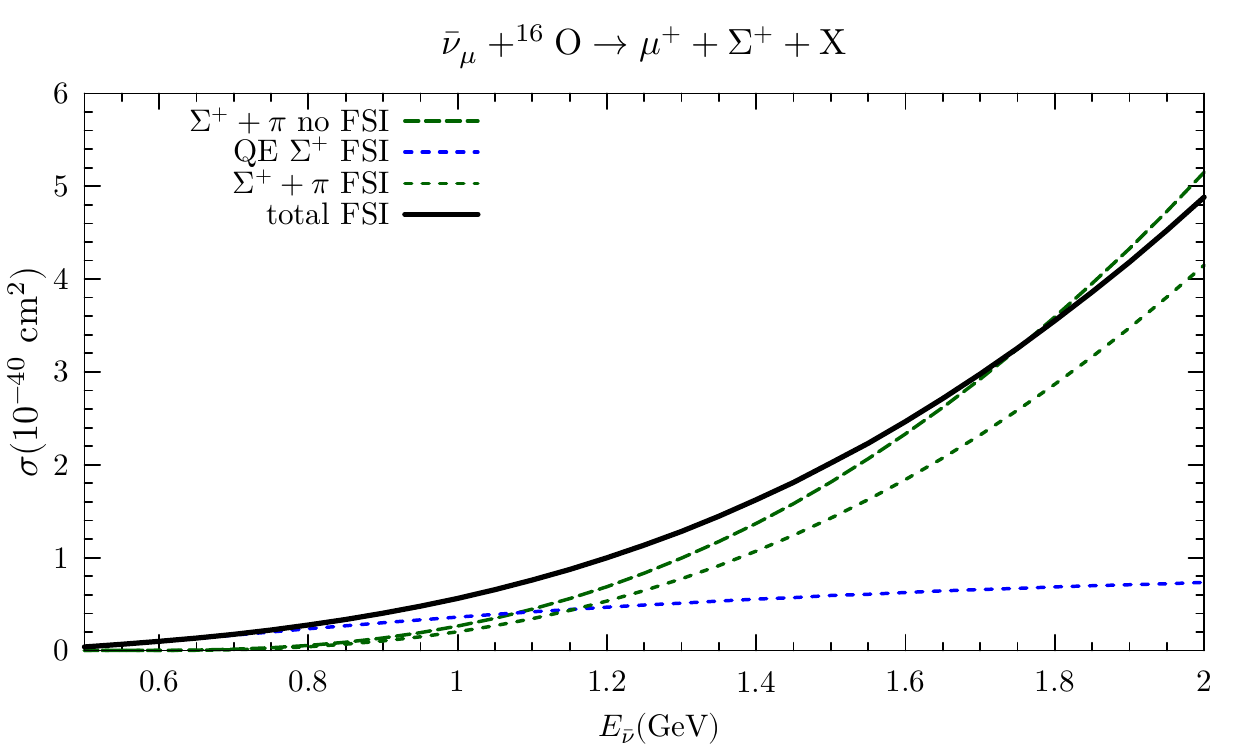}
\hspace{0.4cm}
\centering
\includegraphics[width=0.46\textwidth]{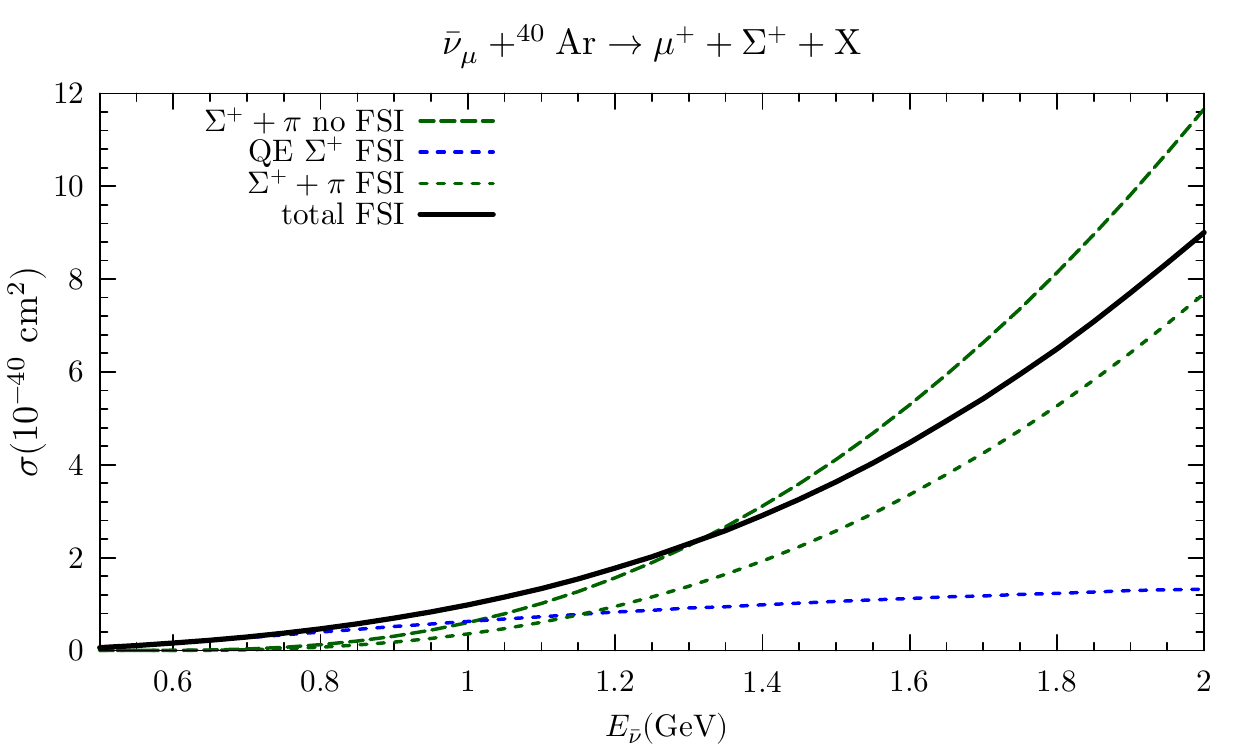}
\caption{Integrated cross sections for quasielastic hyperon
  production, inelastic hyperon-pion production and their sum on
  $^{16}$O (left) and $^{ 40}$Ar (right) as a function of the
  Laboratory $\bar{\nu}_\mu$ energy. Results with and without hyperon
  FSI are displayed. Note that $\Sigma^+$ arise from QE collisions
  only following FSI.}
\label{fig:total_xsect}
\end{figure*}

Integrated cross sections on $^{16}$O and $^{40}$Ar as a function of
$E_{\bar\nu}$ are shown in Fig.~\ref{fig:total_xsect} for $\Lambda$
and $\Sigma^{+,-,0}$ in the final state. Results without and with FSI
are displayed.
We observe the clear dominance of $\Lambda$ production. This channel
has a substantially larger cross section than the other ones taken
together.  This is partly due to FSI which favors
$\Sigma\rightarrow\Lambda$ transitions. Thus, we can see that
$\Lambda$ production is slightly enhanced by FSI\footnote{If events
with $\Lambda$ hyperons trapped in the mean field potential were
counted, the $\Lambda$-production cross section would be increased by
about 20\% (30\%) in $^{16}$O ($^{40}$Ar).}  whereas $\Sigma^-$ and
$\Sigma^0$ production are diminished. The case of $\Sigma^+$ is
especial because there is no direct QE production. Therefore, for
$E_{\bar\nu} < 1.2$~GeV the cross section comes predominantly from
FSI.

For both $\Lambda$ and $\Sigma^-$, QE mechanisms are clearly larger
than the $Y\pi$ ones at the examined range of energies. The situation
is different for $\Sigma^0$, where the cross sections are already
comparable at the upper end.  In the $\Sigma^+$ case, $\Sigma^+ \pi$
primary production becomes gradually dominant for $E_{\bar\nu} >
1.2$~GeV. Moreover, in all channels, the higher threshold implies that
the $Y\pi$ mechanisms start to contribute at higher energies but their
contributions grow faster. This is very clearly visible in the
$\Sigma^0$ case.  Overall, the inelastic mechanisms studied here
provide a substantial part of the hyperon production cross section and
neglecting them would lead to a serious underestimation of strangeness
production induced by antineutrinos.

We should recall that the present model for the inelastic $Y\pi$
processes incorporates the decuplet resonance $\Sigma^*(1385)$, which
provides the biggest contribution to $\Lambda\pi$
production~\cite{BenitezGalan:2021jdm}.  However, the model does not
include the $s$-wave $\Lambda(1405)$ resonance. The comparisons
between the results of Ref.~\cite{BenitezGalan:2021jdm} and
\cite{Ren:2015bsa} show that $\Lambda(1405)$ plays a moderate role in
$\Sigma \pi$ production at low
energies~\cite{BenitezGalan:2021jdm}. However, the $\Lambda\pi$
channel is not affected by the $\Lambda(1405)$ resonance, which only
decays to $\Sigma\pi$ and $\bar{K}N$. Furthermore, $\Lambda(1405)$
mechanisms are dominated by $s$-waves and grow rather slowly as a
function of the energy. As a consequence, after the inspection of
Refs.~\cite{BenitezGalan:2021jdm,Ren:2015bsa}, we conclude that they
would only be competitive with the $\Sigma\pi $ inelastic production
mechanisms considered here for the $\Sigma^{-,0}$ and at energies
$E_{\bar{\nu}}<1.4$ GeV, where QE production is anyway much
larger. Thus, we neglect $\Lambda(1405)$ excitation, as its net
contribution would be small while keeping in mind that it could be
relevant for the study of $\Sigma \pi$ exclusive channels where the
pion is detected in coincidence with the hyperon.

\begin{figure}[ht]
\centering
\includegraphics[width=0.45\textwidth,height=.30\textwidth]{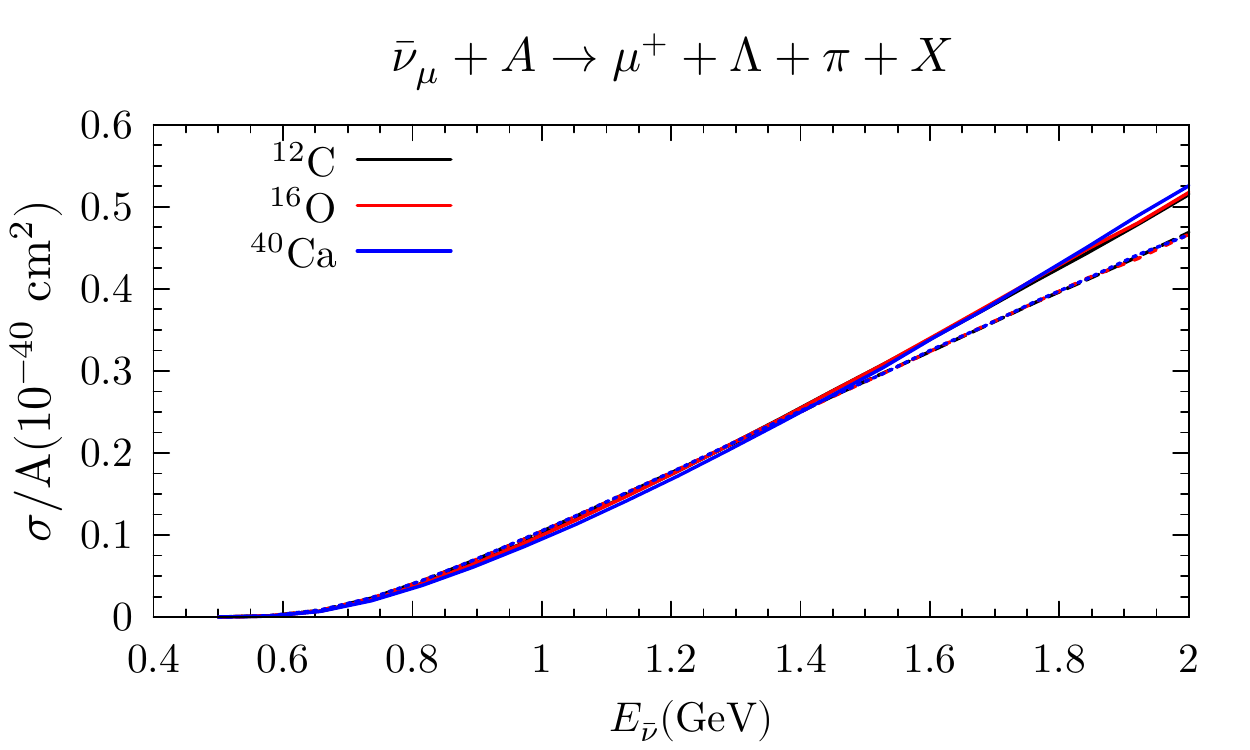}
\centering
\includegraphics[width=0.45\textwidth,height=.30 \textwidth]{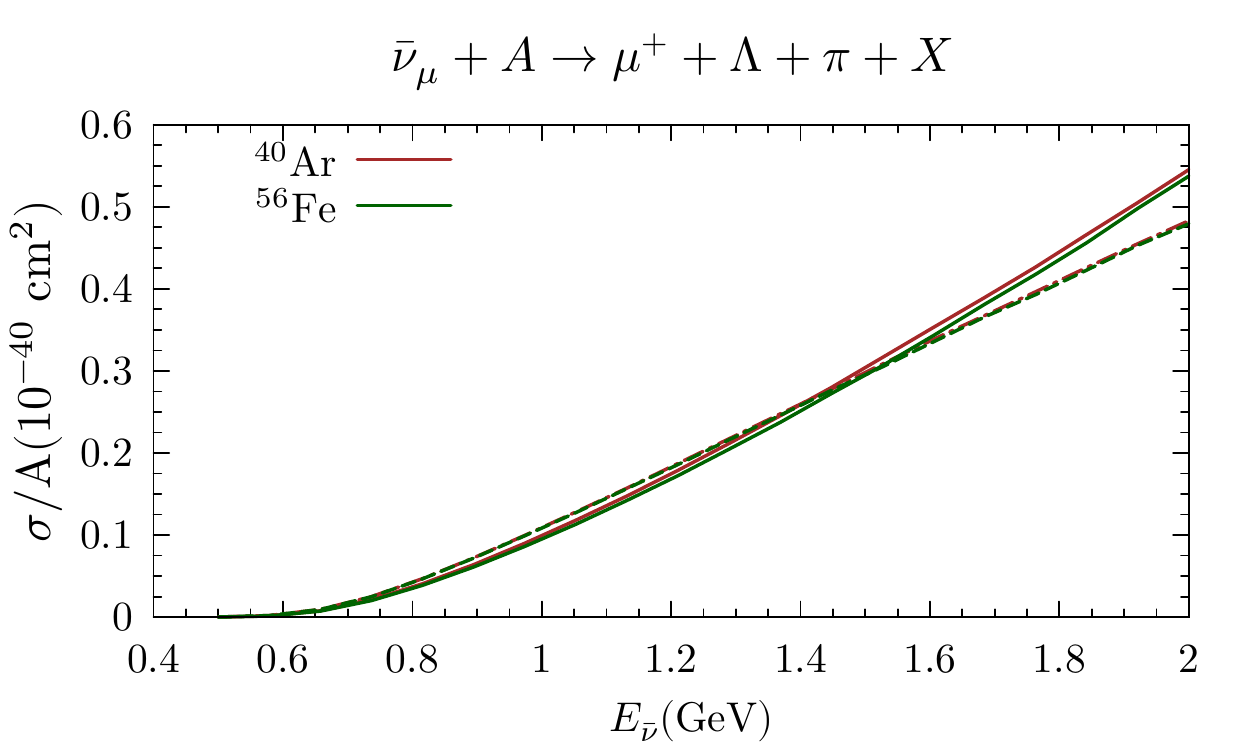}
\caption{Integrated cross sections per nucleons for $\Lambda\pi$
  production as a function of the antineutrino energy for a selection
  of symmetric ($^{12}$C, $^{16}$O and $^{40}$Ca) (top panel) and
  asymmetric nuclei ($^{40}$Ar and $^{56}$Fe) (bottom panel). Dashed
  (solid) lines are obtained without (with) FSI.}
\label{fig:sym_asym}
\end{figure}

In Fig.~\ref{fig:sym_asym}, we compare the integrated cross section
for $\Lambda\pi$ production on symmetric nuclei like ${}^{12}$C,
${}^{16}$O and ${}^{40}$Ca versus two asymmetric ones, ${}^{40}$Ar and
${}^{56}$Fe. To render the comparison clearer, the cross sections are
divided by the mass number. We observe that the cross sections per
nucleon without FSI are almost indistinguishable among nuclei of the
same category (symmetric or asymmetric), but somewhat larger for
asymmetric ones. This is because the cross section for $\Lambda\pi$
production off free neutrons is exactly twice the one off protons, as
shown in Fig. 4 of Ref.~\cite{BenitezGalan:2021jdm}.

When FSI and the $\Lambda$ potential are taken into account, the cross
sections per nucleon get enhanced at upper end of the antineutrino
energy range under consideration. This increase, which is also larger
for asymmetric nuclei is caused by $\Sigma\rightarrow\Lambda$
conversion. In contrast, a very small reduction can be noticed,
particularly for asymmetric nuclei, which can be attributed to the
presence of the attractive potential, on which some of the slow
$\Lambda$ get trapped. It is worth noticing that the scaling with the
number of nucleons is largely preserved in presence of FSI.

We have also estimated the fraction of absorbed pions in the way
explained at the end of the previous section. As expected, it grows
with the nuclear mass number once pions have to travel longer in
heavier nuclei, which increases the absorption probability. On the
other hand, this fraction decreases with the incident antineutrino
energy: the absorption probability increases for slower (in average)
pions. More quantitatively, 55-70\% of the produced pions are absorbed
at $E_{\bar\nu}=1$~GeV and 30-50\% at 2~GeV for the targets considered
in this study (${}^{12}$C, ${}^{16}$O, ${}^{40}$Ca, ${}^{40}$Ar and
${}^{56}$Fe).

\subsection{Hyperons energy distributions}\label{subsec:hyperon}

\begin{figure*}[!ht]
\centering
\includegraphics[width=0.46\textwidth]{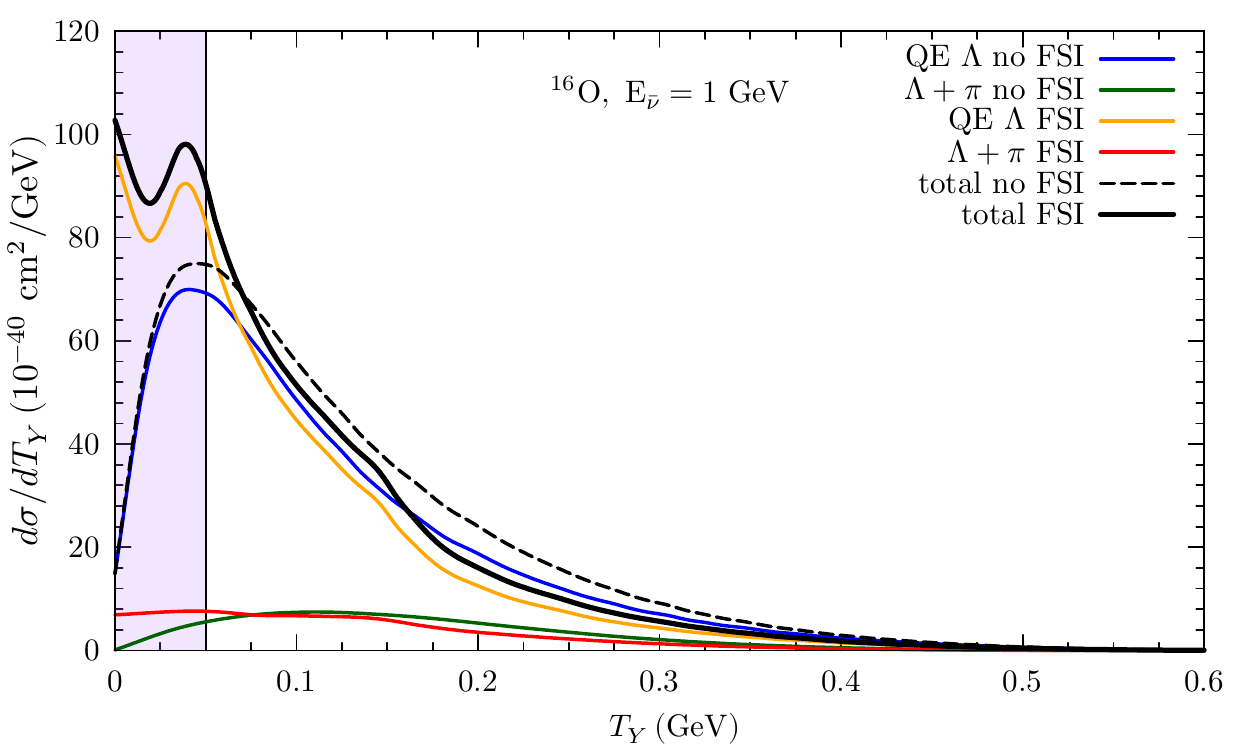}
\hspace{0.4cm}
\centering
\includegraphics[width=0.46\textwidth]{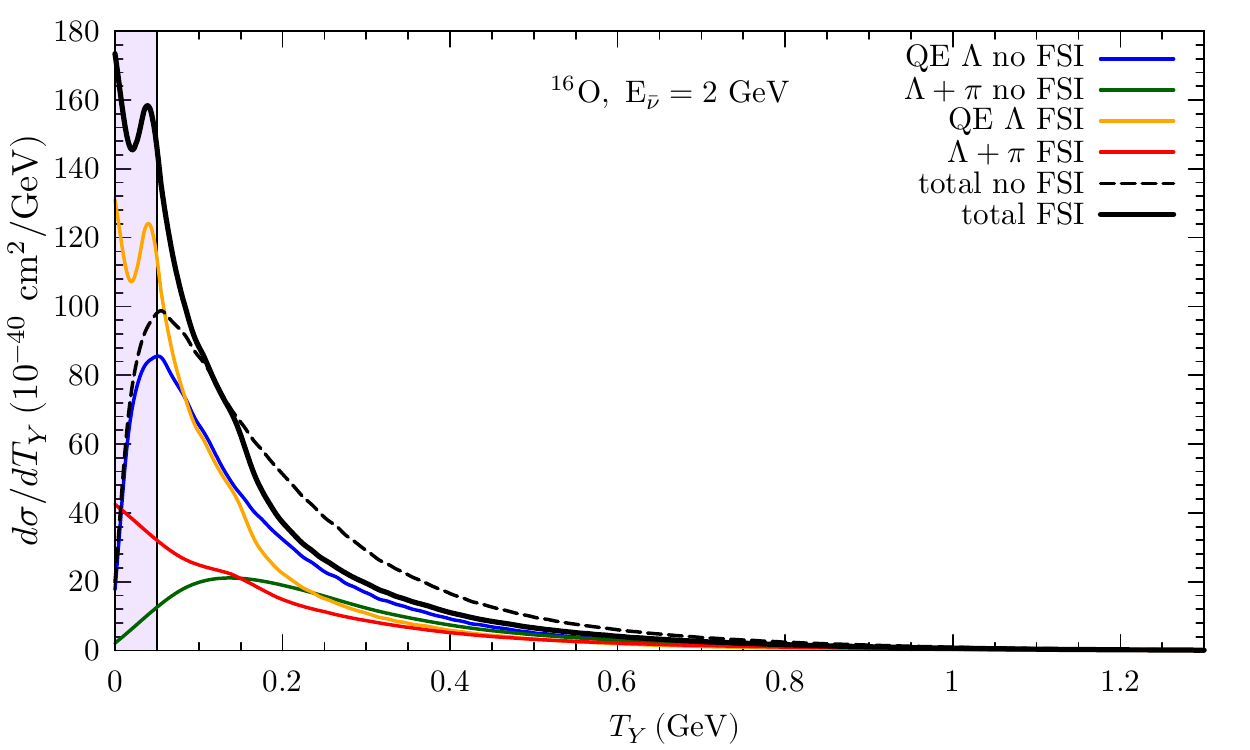}
\centering
\includegraphics[width=0.46\textwidth]{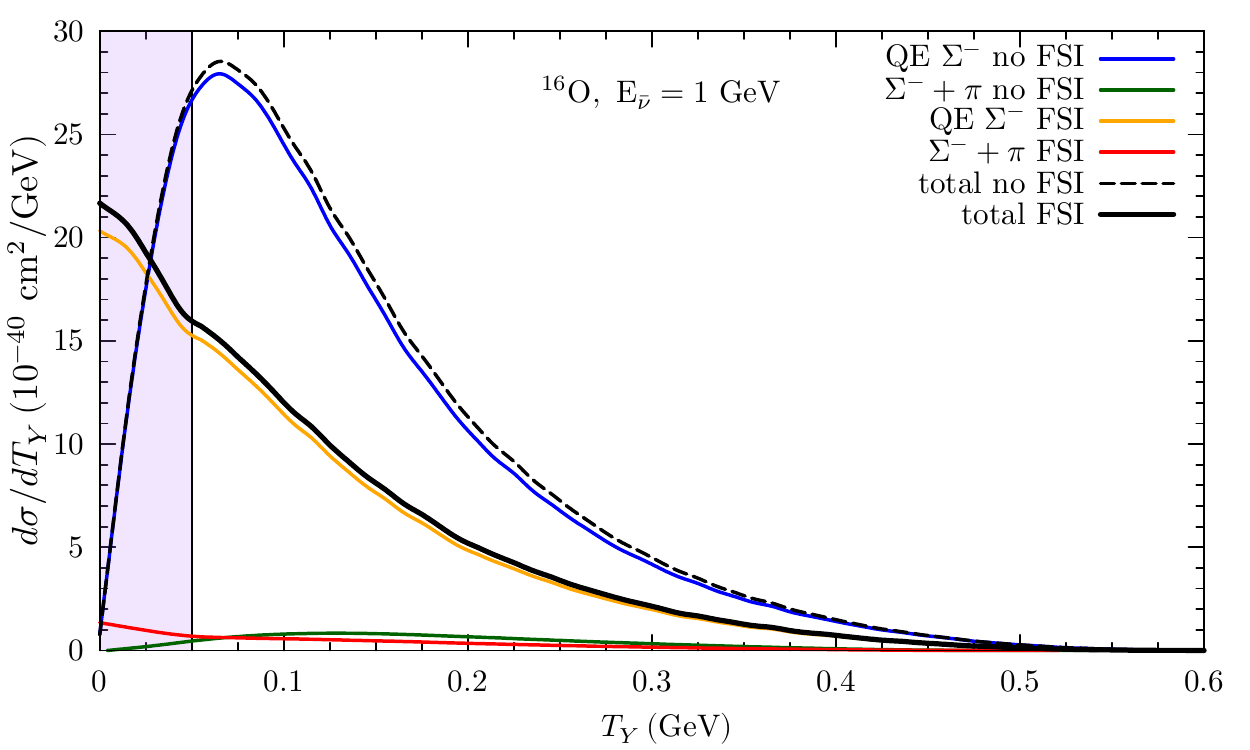}
\hspace{0.4cm}
\centering
\includegraphics[width=0.46\textwidth]{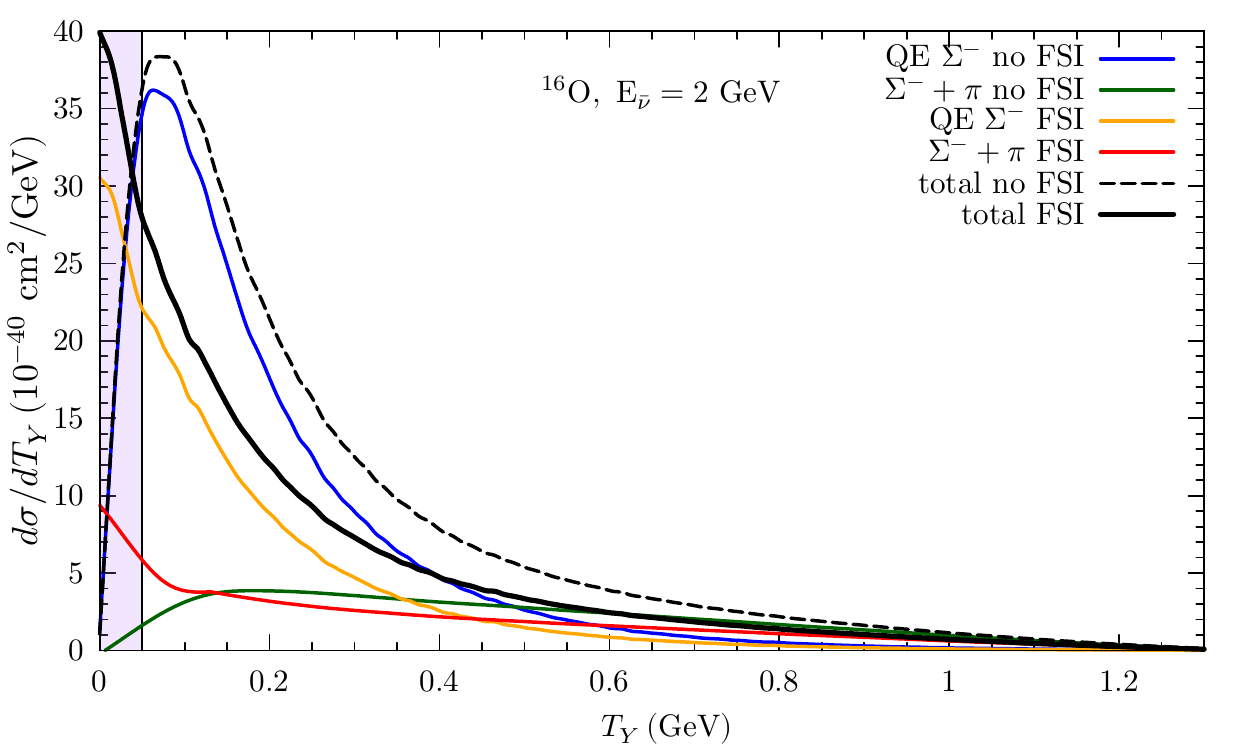}
\centering
\includegraphics[width=0.46\textwidth]{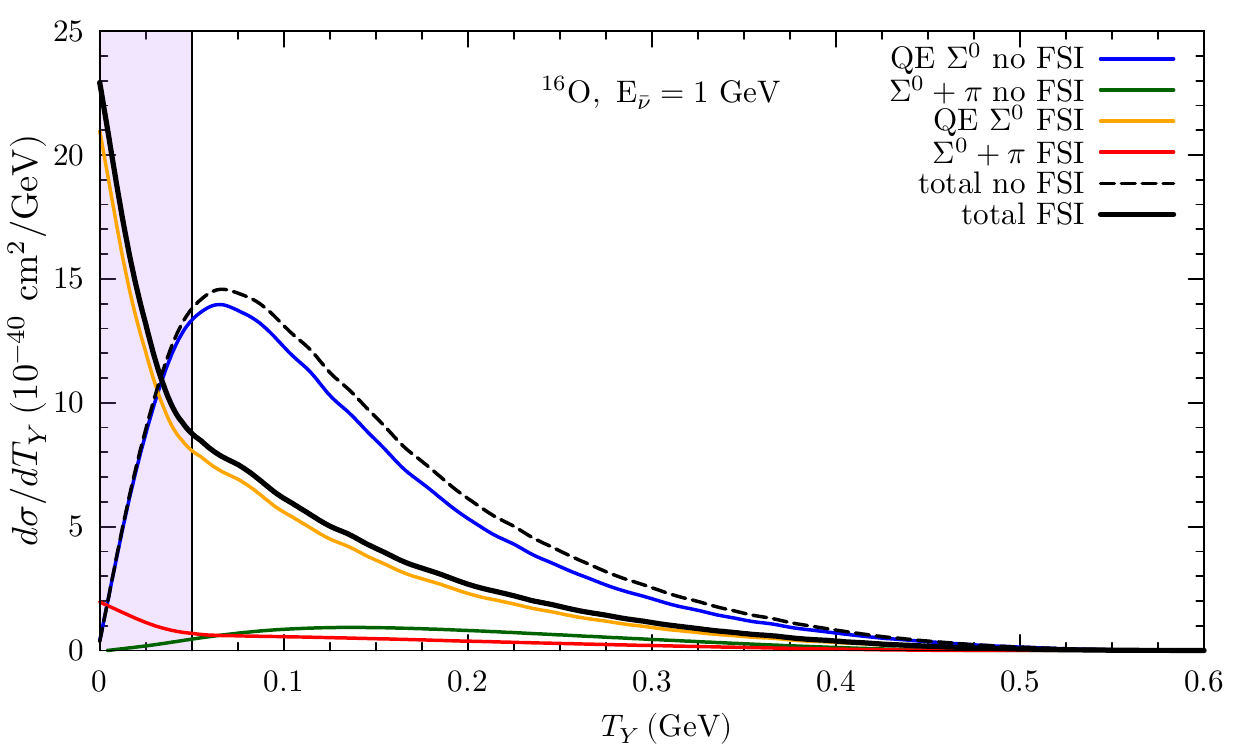}
\hspace{0.4cm}
\centering
\includegraphics[width=0.46\textwidth]{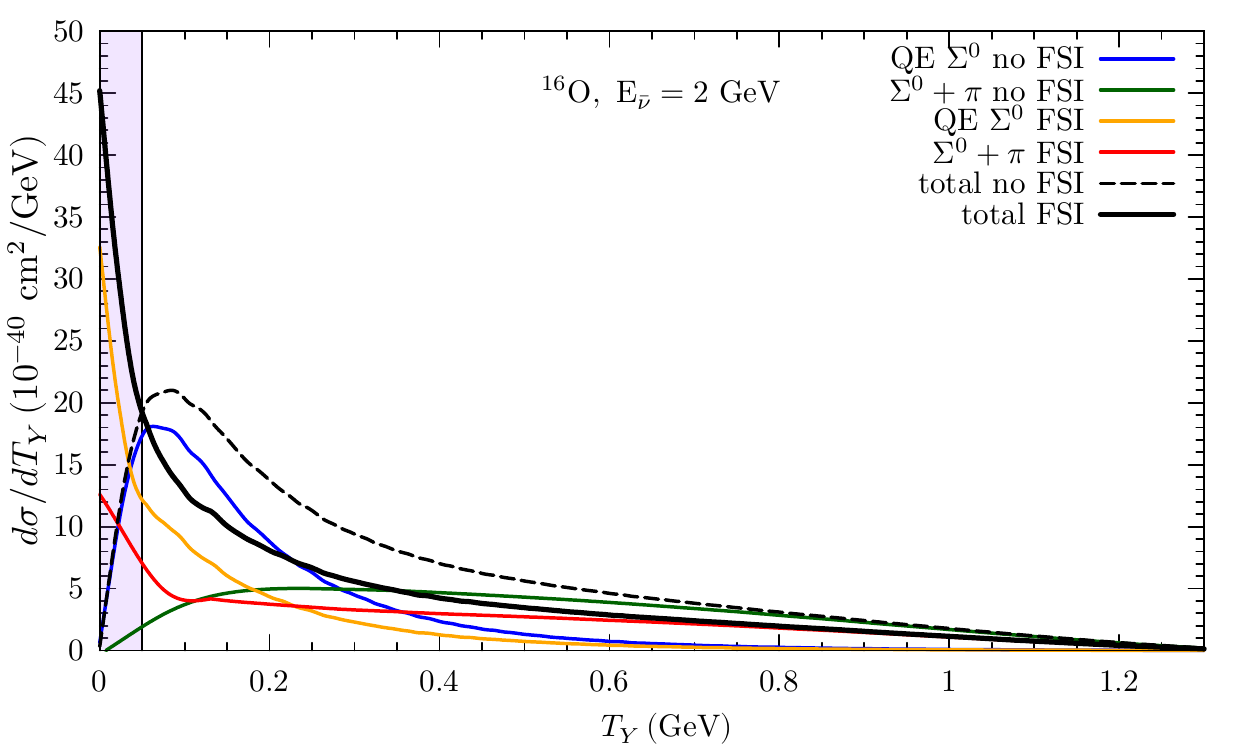}
\centering
\includegraphics[width=0.46\textwidth]{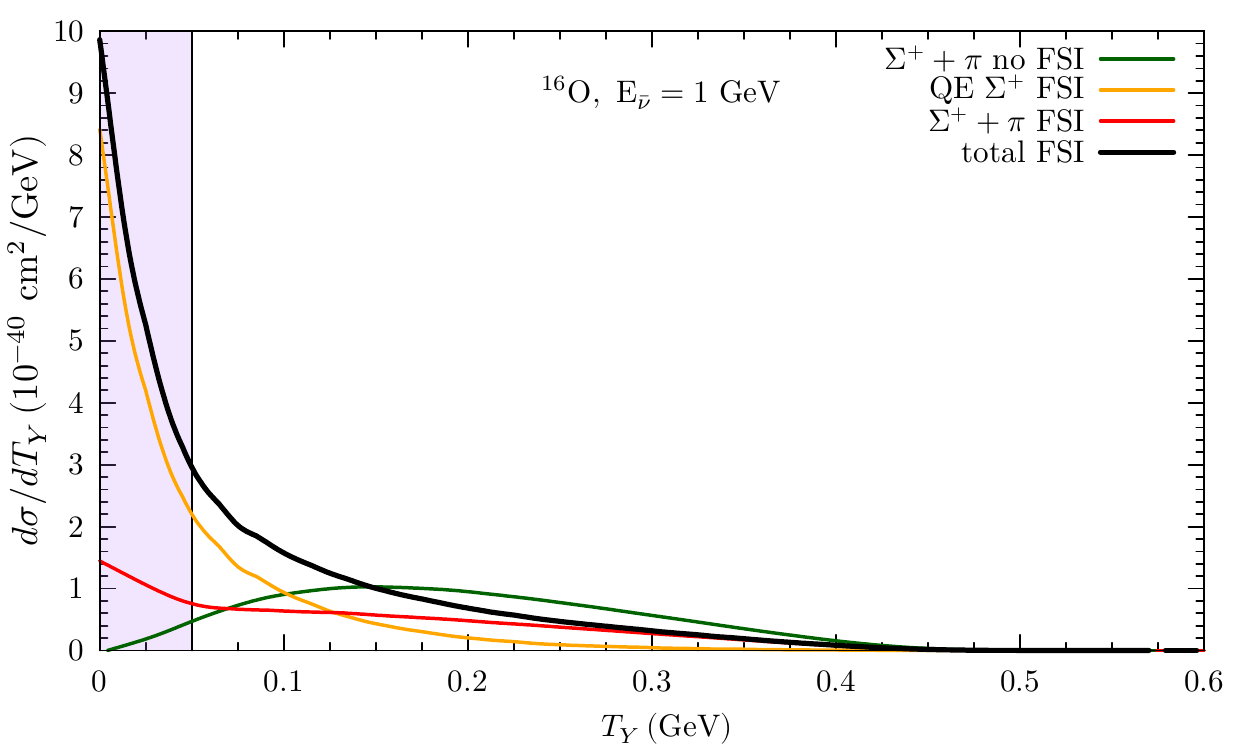}
\hspace{0.4cm}
\centering
\includegraphics[width=0.46\textwidth]{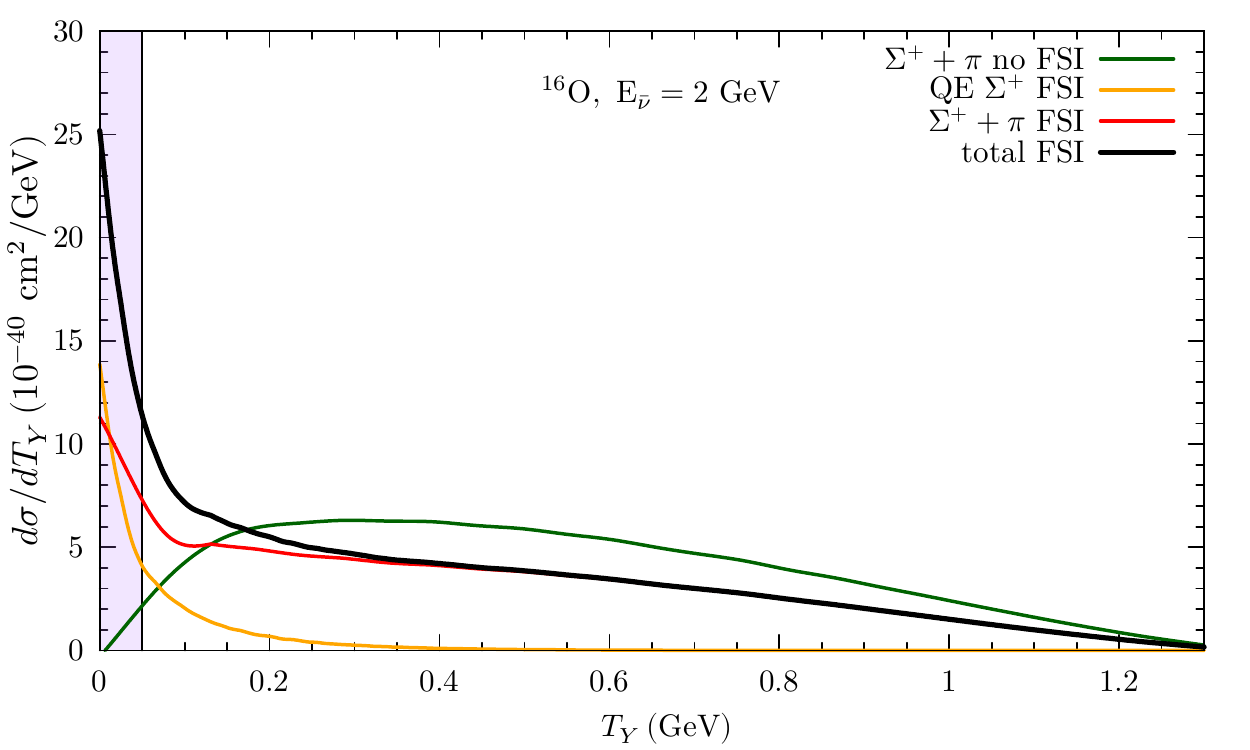}
\caption{Hyperon kinetic energy distributions for quasielastic hyperon
  production, inelastic hyperon-pion production and their sum ,
  computed for $\bar{\nu}_\mu + {}^{16}\rm{O} \rightarrow \mu^+ + Y +
  X$, with and without FSI, at fixed antineutrino energies of
  $E_{\Bar{\nu}}$ = 1 GeV (left panels) and $E_{\Bar{\nu}}$ = 2 GeV
  (right panels). The shaded areas correspond to $T_Y\leq$ 50 MeV.}
\label{fig:energy_distrib_16O}
\end{figure*}

In Fig. \ref{fig:energy_distrib_16O}, we show hyperon kinetic energy
distributions for $^{16}$O at $E_{\Bar{\nu}}$ = 1, 2 GeV.  The
incoming antineutrino energy of 1~GeV and target have been chosen to
facilitate the comparison of our results with previous studies of QE
hyperon production~\cite{Singh:2006xp,Sobczyk:2019uej}. The higher
energy, 2 GeV, examines a kinematic region where $Y\pi$ mechanisms are
more relevant. The shaded bands in Fig.~\ref{fig:energy_distrib_16O}
correspond to hyperon kinetic energies $T_Y\leq$ 50 MeV. As previously
discussed, for hyperon low energies the semiclassical approximation
adopted for FSI is questionable. Even if less accurate, we show
results in this low energy region as they reflect the amount of
produced hyperons and the impact of the $\Lambda$ potential and FSI.

For the QE mechanism, in the absence of the $\Lambda$ potential, we
have checked that our results are fully consistent with those of
Ref.~\cite{Sobczyk:2019uej}\footnote{There are some discrepancies with
the FSI curves of Ref.~\cite{Singh:2006xp} because of a wrong
implementation of the Pauli blocking which led to an underestimation
of FSI effects in that study.}.
As it was also seen in Fig.~\ref{fig:total_xsect}, at 1 GeV QE
contributions are much larger than $Y\pi$ production except for the
$\Sigma^+$ channel, where primary QE production is absent. In this
case, for hyperon kinetic energies just above 100 MeV, inelastic
mechanisms compete with QE $\Sigma^+$ production. In the other three
channels changes caused by the $Y\pi$ mechanisms are minor.

At 2 GeV the relative importance of the $Y\pi$ mechanisms is
larger. They become dominant for $\Sigma^+$ production except at low
$T_Y$. Even for the $\Sigma^0$ they are the main contribution above
300 MeV.

This figure clearly illustrates the consequences of FSI. Not only many
of the hyperons mutate into another species. There is also a clear
shape distortion caused by a significant event displacement towards
low kinetic energies because in each interaction hyperons transfer a
fraction of their energy to the scattered nucleon. In fact, the energy
distributions in presence of FSI are peaked at low energies.  This
feature is most important for $\Sigma^+$, a large fraction of which
are emitted after FSI. In the case of $\Lambda$ production, the impact
of the potential is significant at low energies, where it leads to
changes in the energy spectra. After FSI, many $\Lambda$ hyperons have
kinetic energies smaller than the absolute value of the local mean
field potential and therefore, as explained in Sec.~\ref{sec:fsi}, are
not counted as asymptotic states. We do not dwell on the details of
the shape of $\Lambda$ kinetic energy distributions at $T_\Lambda \leq
50$~MeV, also obtained in Ref.~\cite{Thorpe:2020tym}, because, as
explained, the semiclassical treatment is not realistic at these
energies.

\subsection{Angular distributions}
\label{sec:angle}
\begin{figure*}[ht]
\centering
\includegraphics[width=0.49\textwidth,height=.30\textwidth]{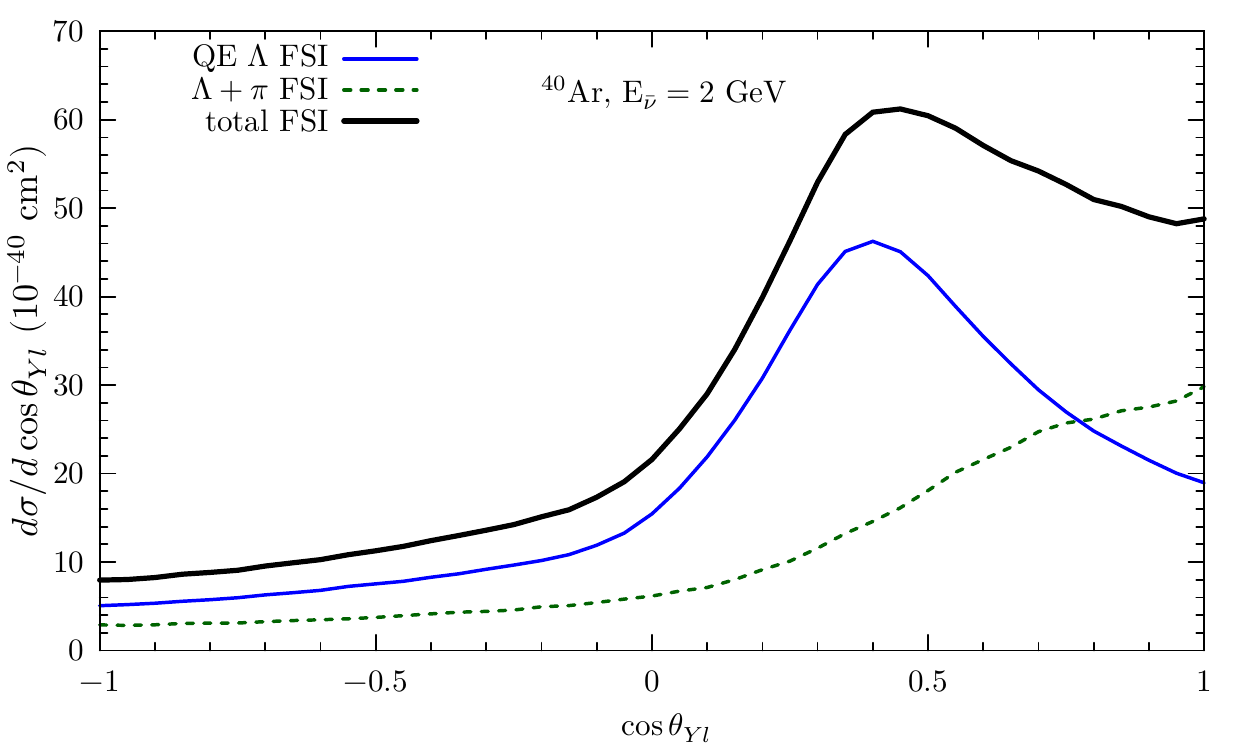}
\centering
\includegraphics[width=0.49\textwidth,height=.30\textwidth]{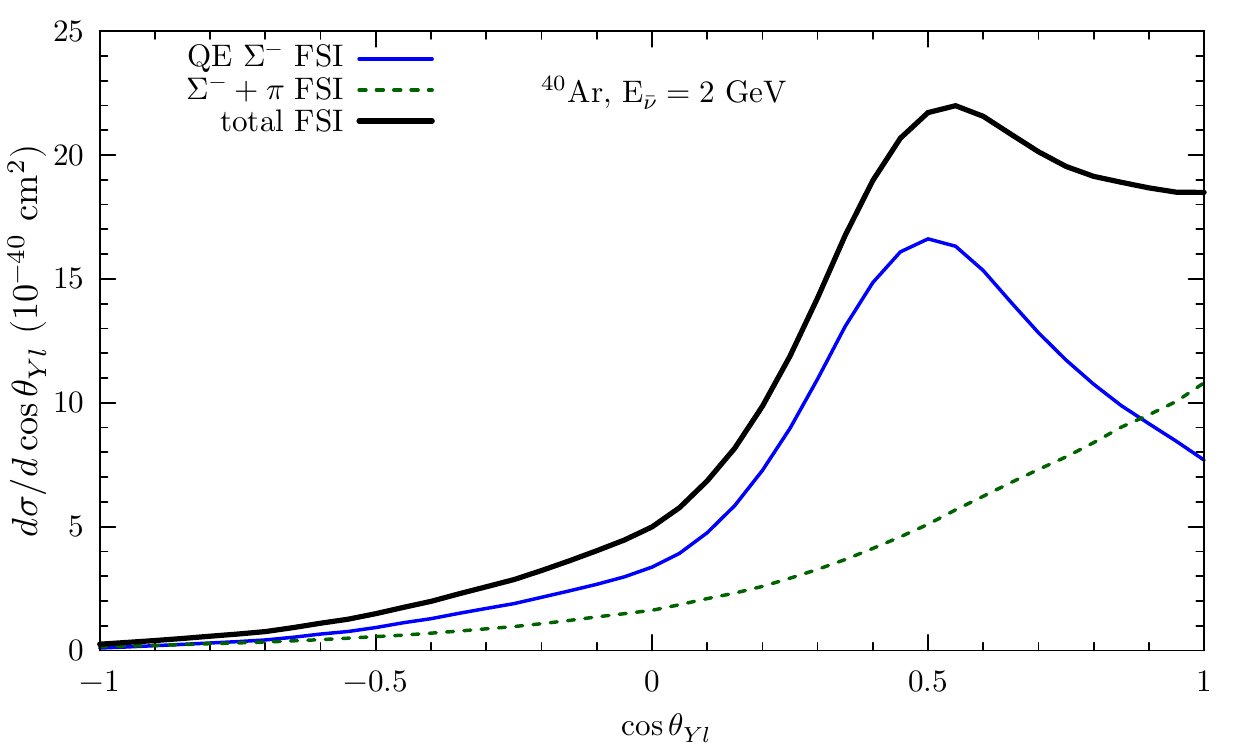}
\centering
\includegraphics[width=0.49\textwidth,height=.30\textwidth]{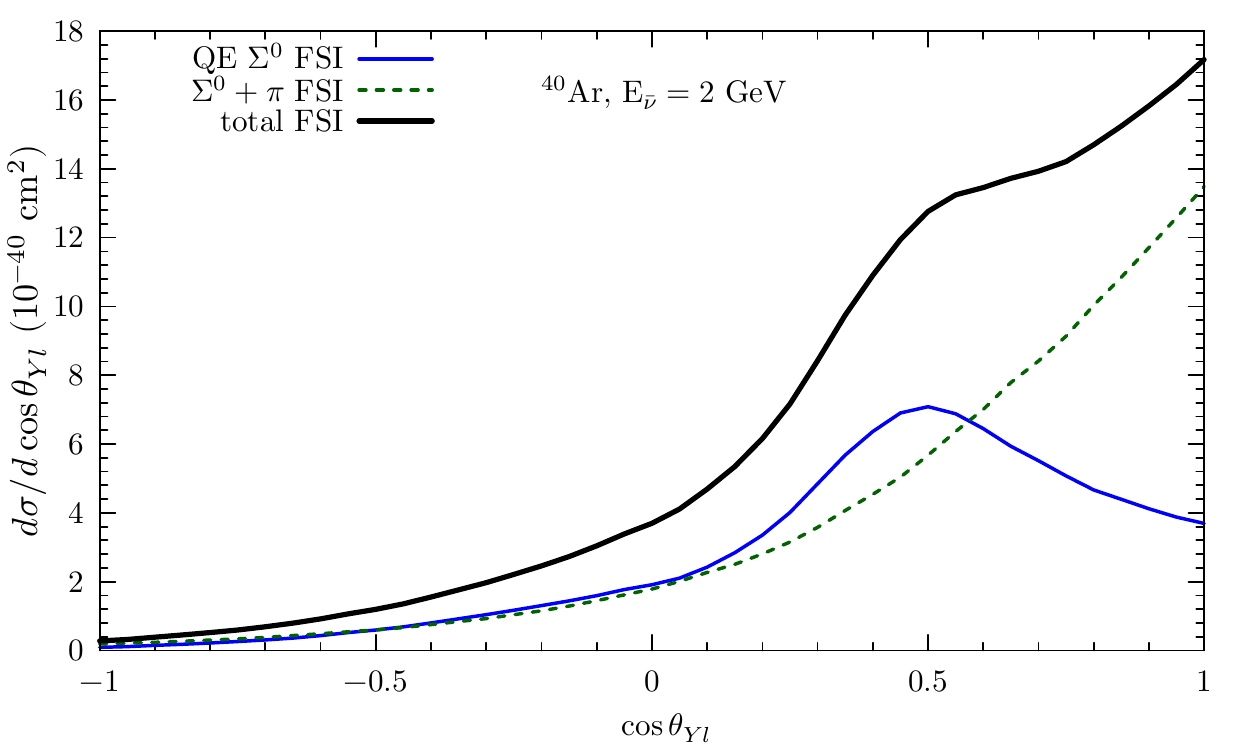}
\centering
\includegraphics[width=0.49\textwidth,height=.30\textwidth]{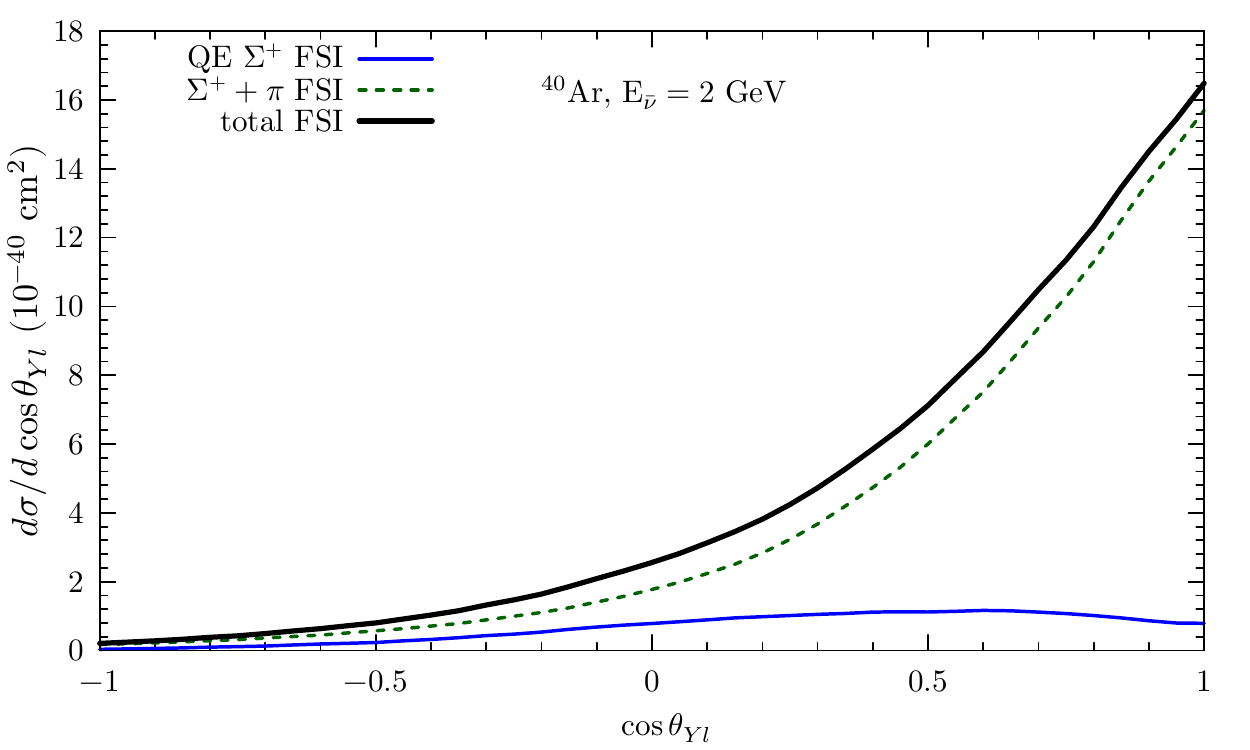}
\caption{Angular distributions for the cosine of the angle between
  hyperon and muon in $\bar{\nu}_{\mu}+{}^{40}\mathrm{Ar}\rightarrow
  \mu^++Y+X$.  Solid blue lines correspond to QE reactions, dashed
  lines to $Y\pi$ production, and solid black ones to their sum. All
  the curves include FSI and the $\Lambda$ nuclear potential.}
\label{fig:angles}
\end{figure*}
We have examined angular distributions for the relative angle between
final hyperon and lepton momenta, $\theta_{Yl}$.\footnote{Pions are
treated inclusively: they might be emitted or absorbed.}  We have
found that $d\sigma/d\cos \theta_{Yl}$ is particularly sensitive to
the production mechanisms, showing a different behavior for QE and
$Y\pi$ processes. This distinctive feature could help to disentangle
them using data and study their relative importance.

Some illustrative results are presented in Fig.~\ref{fig:angles}.  For
all channels, the $Y\pi$ contributions are forward peaked and display
a monotonous growth as a function of $\cos \theta_{Yl}$. On the other
hand, for QE processes there is a bump around $\cos \theta_{Yl}\approx
0.4 $, which is also present in the total cross section. This peak,
mostly driven by phase space, exists for QE processes on single
nucleons. It is also present when vector and axial parts of the
hadronic current are separately considered. The peak position hardly
changes with the neutrino energy and is largely unaffected by FSI. The
presence of such a structure for the QE mechanism and its properties
are in line with the findings of Ref.~\cite{Thorpe:2020tym} as can be
seen in Figs.~23 and 24 of that reference. As shown in
Fig.~\ref{fig:angles} up to $E_{\bar\nu} = 2$~GeV the peak remains
visible in the sum of QE and $Y\pi$ for all hyperons except for
$\Sigma^+$ because the QE contribution is too small.

\subsection{Comparison with the MicroBooNE measurement}
\label{sec:MicroBooNE}
As discussed in the Introduction, the experimental information for
weak hyperon production is very scarce but a new scenario awaits with
SBND at FNAL, which is expected to accumulate 8000 $\Lambda$ and 4500
$\Sigma^+$ in only three years of
operation~\cite{Brailsford:2017rxe,Machado:2019oxb}\footnote{This
estimate was obtained by the SBND collaboration using the GENIE event
generator~\cite{Andreopoulos:2009rq} for a $6.6\times10^{20}$ protons
on target exposure.}.

In the mean time, MicroBooNE has reported five $\Lambda$ events over
the background from the exposure of its liquid argon time projection
chamber to the off-axis NUMI beam~\cite{MicroBooNE:2022bpw}, and a
fourfold increase with already collected data is
expected~\cite{MicroBooNE:2022bpw}. Driven by these prospects we have
studied $\Lambda$ production on argon in the conditions of the
MicroBooNE measurement.  Clearly, the statistics is still too low to
discriminate between models or to attempt the extraction of transition
form factors or other parameters of the theory but the available data
already provide useful information.

\begin{figure*}[!ht]
\centering
\includegraphics[width=0.48\textwidth,height=.30\textwidth]{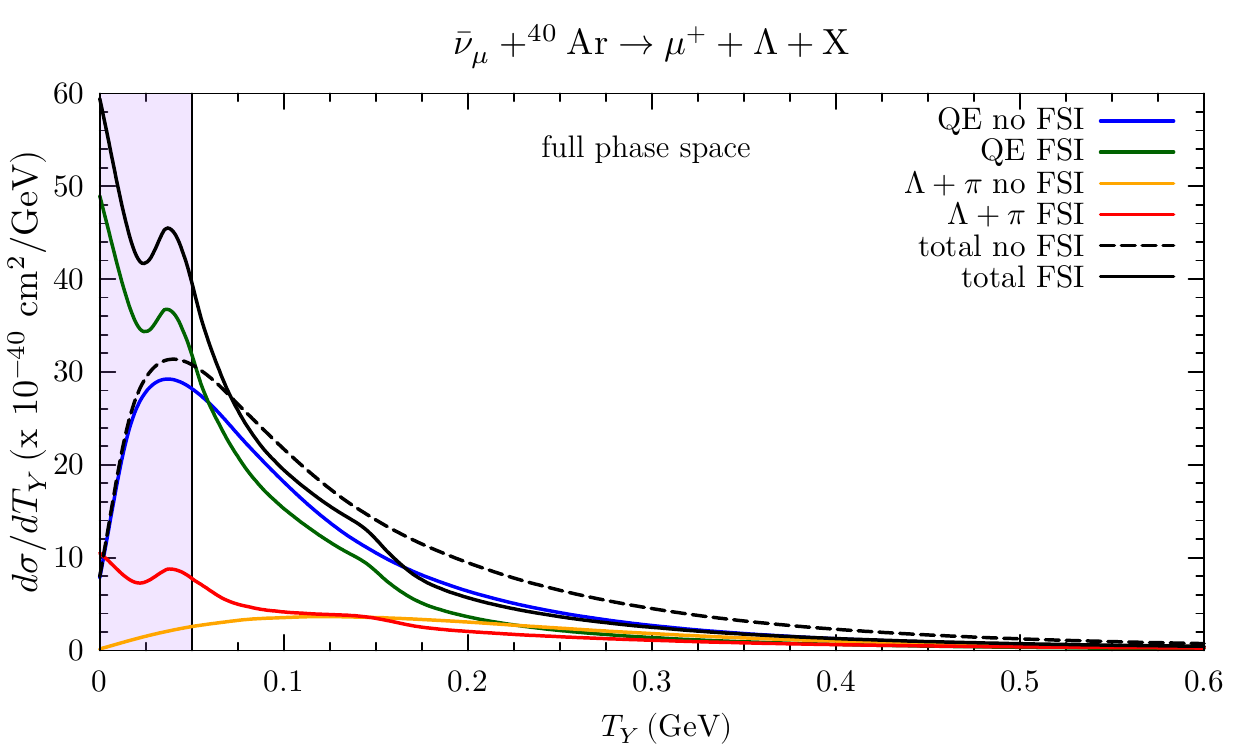}
\includegraphics[width=0.45\textwidth,height=.30\textwidth]{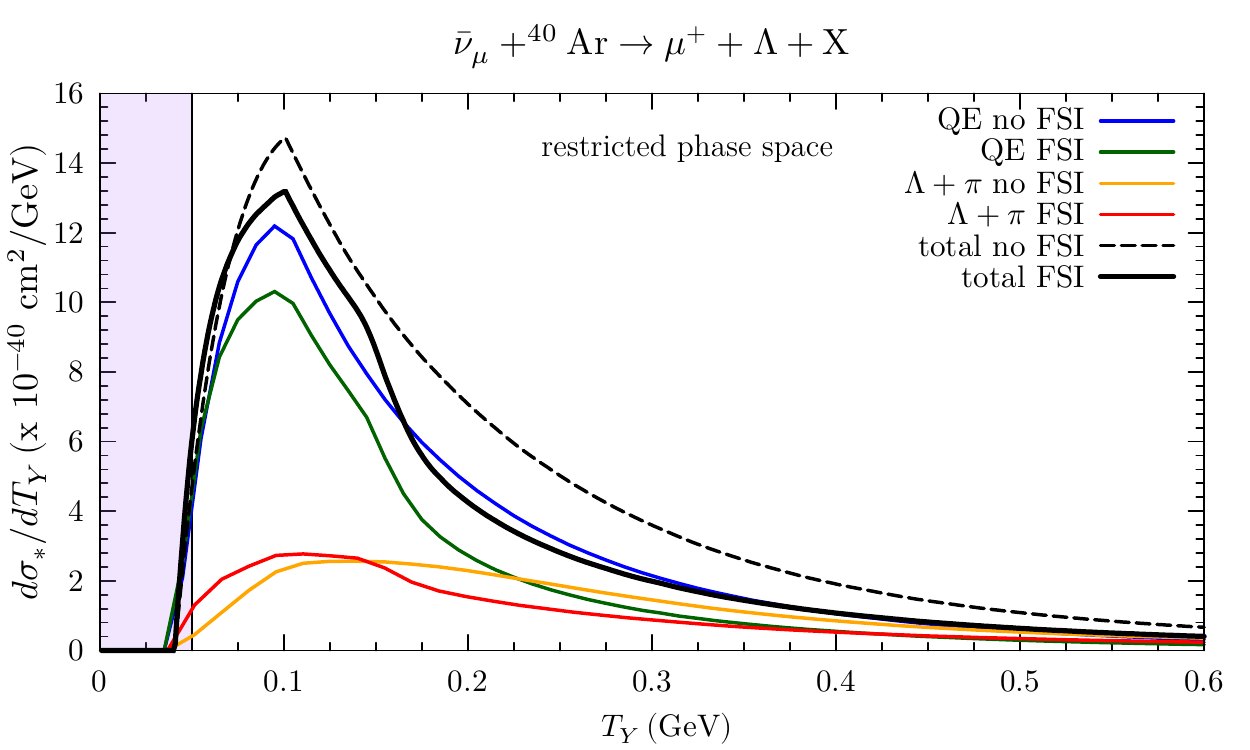}
\caption{$\Lambda$ kinetic energy spectra averaged over the NUMI flux
  at MicroBooNE without (left) and with (right) experimental phase
  space restrictions~\cite{MicroBooNE:2022bpw}. Contributions from QE
  and $Y\pi$ mechanisms and their sum, with and without FSI, are
  shown.}
\label{fig:microbooneflux}
\end{figure*}
In Fig.~\ref{fig:microbooneflux} (left), we plot the differential
cross section for $\Lambda$ production as a function of the hyperon
kinetic energy averaged over the flux used by MicroBooNE in their
simulations. Its shape is available from the Supplemental Material of
Ref.~\cite{MicroBooNE:2022bpw}. With respect to the distribution
without FSI, a clear enhancement at low kinetic energies is apparent
which reveals a strong $\Sigma \to \Lambda$ conversion. The QE
contribution is predominant while the $Y\pi$ one brings only a minor
increase.  According to our estimates based on Eq.~\ref{eq:eikonal},
half of the primarily produced pions will be absorbed and the rest are
part of the final hadronic system $X$ without strange particles.
However, the MicroBooNE measurement has phase space restrictions
dictated by the detection thresholds of the $\Lambda \to p \, \pi^-$
decay products used to identify the hyperon. To correct for this we
multiply our prediction by the fraction of $\Lambda$ decays with $p$
and $\pi^-$ above detection threshold. This quantity as a function of
the hyperon momentum is readily provided by MicroBooNE in the
Supplemental Material of Ref.~\cite{MicroBooNE:2022bpw}. The result is
displayed in the right panel of Fig.~\ref{fig:microbooneflux}. One
immediately notices that the detector is blind to $\Lambda$ with
$T_\Lambda < 40$~MeV, accounting for a large fraction of the cross
section. The opportunity to understand better FSI and test models in
this challenging region is unfortunately missed.  Interestingly, this
physics will be enabled in large-volume pressurized argon time
projection chambers such as the one under development by
DUNE~\cite{DUNE:2021tad,DUNE:2022yni}, where reconstruction thresholds
in the few-MeV range are anticipated for hadrons.

On the other hand, in the restricted phase space, the relative
importance of the $Y\pi$ mechanism as a source of pions is
considerably enhanced. In fact, as can be seen in
Table~\ref{tab:totalcrosssections}, the $Y\pi$ contribution accounts
to one third of the total flux-averaged cross section in the
restricted phase space, $\sigma_*$.
\begin{table}[h]
    \centering
    \begin{tabular}{|c|c|}
    \hline
         &  $\sigma_*$ ($\times 10^{-40}$~cm$^2$/Ar) \\
    \hline
       MicroBooNE  & $2.0^{+2.1}_{-1.6}$\\
      \hline
       QE + $Y\pi$, full model & $2.13$ 
       \\
      \hline
        QE  & $1.44$ \\
       \hline
        $Y\pi$ & $0.69$ 
        \\
       \hline
    \end{tabular}
    \caption{Flux-averaged cross section for $\Lambda$ production on
      argon in the restricted phase space, $\sigma_*$ defined in
      Ref.~\cite{MicroBooNE:2022bpw}. The experimental value measured
      by MicroBooNE is compared to our result, for which the QE and
      $Y\pi$ contributions are separately given.}
    \label{tab:totalcrosssections}
\end{table}
The value of $\sigma_*$ we obtain, including both QE and $Y\pi$
contributions, agrees with the MicroBooNE experimental result, whose
discriminating power is still limited by the low statistics. According
to Ref.~\cite{MicroBooNE:2022bpw}, the measured $\sigma_*$ value is
also consistent with the predictions from the
GENIE~\cite{Andreopoulos:2009rq} and NuWro~\cite{Thorpe:2020tym} event
generators. It should be however remarked that neither of these two
event generators includes $Y\pi$ channels\footnote{In addition, GENIE
does not take FSI into account.}. Our findings imply that the $\Lambda
\pi$ contribution, which is dominated by $\Sigma^*(1385)$ excitation,
is a very important ingredient for the analysis and interpretation of
the experimental results awaited at MicroBooNE and SBND.

\section{Summary and conclusions}\label{sec:conclusions}

We have studied hyperon production on nuclei induced by antineutrinos,
in the energy region where associated strangeness production ($\Delta
S=0$) and secondary hyperon production induced by $\bar{K}$ are still
negligible.

These reactions proceed mostly via quasielastic scattering in which
the emission of a charged lepton and a hyperon is induced by
antineutrino-nucleon interactions. These mechanisms have been earlier
studied, exploring their sensitivity to the nucleon-to-hyperon
transition form factors, SU(3) symmetry breaking or the existence of
second class currents.
  
In this work, we have also investigated an additional type of
processes which contributes to the inclusive hyperon production cross
section. Namely, the case in which a pion is emitted together with a
$\Sigma$ or $\Lambda$. The extra pion mass implies a higher threshold
than in the QE mechanisms but, nevertheless, these $Y\pi$ processes
start to be relevant at much lower energies than $Y K$ production and
their cross sections grow faster with the antineutrino energy than the
QE ones. Moreover, the $\Sigma^*(1385)$, which decays strongly into
$\Lambda\pi$ and $\Sigma\pi$, is located close to threshold.

Nuclear effects have been taken into account using the impulse
approximation and a local Fermi gas description of the initial state.
An attractive nuclear mean field potential for $\Lambda$ hyperons has
also been incorporated.  The final state interaction between the
produced hyperon and the nucleon spectators has been accounted for
with a Monte Carlo intranuclear cascade.  As a consequence of hyperon
FSI, there is an enhancement in $\Lambda$ production caused by $\Sigma
\to \Lambda$ conversion. In addition, primary produced hyperons lose
energy when they collide with nucleons and, thus, energy distributions
are strongly shifted towards low energies.  Instead, lepton-hyperon
angular distributions are only slightly softened by FSI.

We find that $Y\pi$ channels are relevant for the description of the
hyperon production off nuclei. Indeed, they provide the main
contribution to $\Sigma^+$ production and generate a sizable fraction
of the total cross section in other channels. Their relative
importance increases with energy.  Failure to account for these
mechanisms would introduce biases in the experimental analysis and
interpretation of experimental results. This will be the case, for
example, in attempts to constrain nucleon-to-hyperon transition form
factors or extract information about hyperon potentials using neutrino
scattering. $Y\pi$ events could be discriminated by detecting the
emitted pions but this will not be possible for a large fraction of
these events because of pion intranuclear absorption and detection
thresholds.  In this context we have obtained that distributions over
the lepton-hyperon relative angles are useful observables for
distinguishing between QE and $Y\pi$ processes. In any respect,
adequate consideration of inelastic $Y\pi$ production in the event
generators used by experiments, such as GENIE or NuWro, would be
required.

Finally, we have studied $\Lambda$ production on argon in the
conditions of the recent MicroBooNE measurement. This implies folding
with the antineutrino flux and imposing the proper acceptance cuts.
The relative high detection threshold and small acceptance for low
energy $\Lambda$ strongly reduces the fraction of events that can be
identified, and increases the relative importance of inelastic $Y\pi$
with respect to QE ones. In our result for the phase space restricted
flux averaged cross section, which is consistent with the
low-statistics experimental value, the $\Lambda \pi$ mechanism
accounts for one third of the total.

More data on hyperon production from MicroBooNE and SBND are eagerly
awaited to learn more about this rare but interesting process.

\section{Acknowledgements}\label{sec:acknowledgements}

The authors acknowledge the use of resources from the scientific
computing cloud PROTEUS of the Instituto Interuniversitario Carlos I
de F\'isica Te\'orica y Computacional of the University of Granada to
perform some of the numerical calculations required in this study.
They also thank J. Nowak and C. Thorpe for very useful communications
regarding hyperon production measurements at MicroBooNE, and
D. González-Díaz for valuable insight about argon time projection
chambers. The authors are also grateful to Prof. J. T. Sobczyk for
other useful information.

This work has been partially supported by the Spanish Ministry of
Science under grants PID2020-114767GB-I00 and PID2020-112777GB-I00,
funded by MCIN/AEI/10.13039/501100011033. It has also been funded by
FEDER/Junta de Andalucia-Consejeria de Transformacion Economica,
Industria, Conocimiento y Universidades/A-FQM-390-UGR20, by the Junta
de Andalucia (grant No.~FQM-225) and by Generalitat Valenciana under
contract PROMETEO/2020/023 and the ``Planes Complementarios de I+D+i"
program (grant ASFAE/2022/022) by MCIN with funding from the European
Union NextGenerationEU and Generalitat Valenciana.  M.B.G. also
acknowledges support from Spanish Ministry of Science under grant
PRE2018-083794 funded by MCIN/AEI/10.13039/501100011033 and by ``ESF
Investing in your future".

\bibliography{biblio}

\begin{thebibliography}{62}%
\makeatletter
\providecommand \@ifxundefined [1]{%
 \@ifx{#1\undefined}
}%
\providecommand \@ifnum [1]{%
 \ifnum #1\expandafter \@firstoftwo
 \else \expandafter \@secondoftwo
 \fi
}%
\providecommand \@ifx [1]{%
 \ifx #1\expandafter \@firstoftwo
 \else \expandafter \@secondoftwo
 \fi
}%
\providecommand \natexlab [1]{#1}%
\providecommand \enquote  [1]{``#1''}%
\providecommand \bibnamefont  [1]{#1}%
\providecommand \bibfnamefont [1]{#1}%
\providecommand \citenamefont [1]{#1}%
\providecommand \href@noop [0]{\@secondoftwo}%
\providecommand \href [0]{\begingroup \@sanitize@url \@href}%
\providecommand \@href[1]{\@@startlink{#1}\@@href}%
\providecommand \@@href[1]{\endgroup#1\@@endlink}%
\providecommand \@sanitize@url [0]{\catcode `\\12\catcode `\$12\catcode
  `\&12\catcode `\#12\catcode `\^12\catcode `\_12\catcode `\%12\relax}%
\providecommand \@@startlink[1]{}%
\providecommand \@@endlink[0]{}%
\providecommand \url  [0]{\begingroup\@sanitize@url \@url }%
\providecommand \@url [1]{\endgroup\@href {#1}{\urlprefix }}%
\providecommand \urlprefix  [0]{URL }%
\providecommand \Eprint [0]{\href }%
\providecommand \doibase [0]{https://doi.org/}%
\providecommand \selectlanguage [0]{\@gobble}%
\providecommand \bibinfo  [0]{\@secondoftwo}%
\providecommand \bibfield  [0]{\@secondoftwo}%
\providecommand \translation [1]{[#1]}%
\providecommand \BibitemOpen [0]{}%
\providecommand \bibitemStop [0]{}%
\providecommand \bibitemNoStop [0]{.\EOS\space}%
\providecommand \EOS [0]{\spacefactor3000\relax}%
\providecommand \BibitemShut  [1]{\csname bibitem#1\endcsname}%
\let\auto@bib@innerbib\@empty
\bibitem [{\citenamefont {Cabibbo}(1963)}]{Cabibbo:1963yz}%
  \BibitemOpen
  \bibfield  {author} {\bibinfo {author} {\bibfnamefont {N.}~\bibnamefont
  {Cabibbo}},\ }\bibfield  {title} {\bibinfo {title} {{Unitary Symmetry and
  Leptonic Decays}},\ }\href {https://doi.org/10.1103/PhysRevLett.10.531}
  {\bibfield  {journal} {\bibinfo  {journal} {Phys. Rev. Lett.}\ }\textbf
  {\bibinfo {volume} {10}},\ \bibinfo {pages} {531} (\bibinfo {year}
  {1963})}\BibitemShut {NoStop}%
\bibitem [{\citenamefont {Workman}\ \emph {et~al.}(2022)\citenamefont {Workman}
  \emph {et~al.}}]{ParticleDataGroup:2022pth}%
  \BibitemOpen
  \bibfield  {author} {\bibinfo {author} {\bibfnamefont {R.~L.}\ \bibnamefont
  {Workman}} \emph {et~al.} (\bibinfo {collaboration} {Particle Data Group}),\
  }\bibfield  {title} {\bibinfo {title} {{Review of Particle Physics}},\ }\href
  {https://doi.org/10.1093/ptep/ptac097} {\bibfield  {journal} {\bibinfo
  {journal} {PTEP}\ }\textbf {\bibinfo {volume} {2022}},\ \bibinfo {pages}
  {083C01} (\bibinfo {year} {2022})}\BibitemShut {NoStop}%
\bibitem [{\citenamefont {Cabibbo}\ and\ \citenamefont
  {Chilton}(1965)}]{Cabibbo:1965zza}%
  \BibitemOpen
  \bibfield  {author} {\bibinfo {author} {\bibfnamefont {N.}~\bibnamefont
  {Cabibbo}}\ and\ \bibinfo {author} {\bibfnamefont {F.}~\bibnamefont
  {Chilton}},\ }\bibfield  {title} {\bibinfo {title} {{Hyperon Production by
  Neutrinos in an $SU_3$ Model}},\ }\href
  {https://doi.org/10.1103/PhysRev.137.B1628} {\bibfield  {journal} {\bibinfo
  {journal} {Phys. Rev.}\ }\textbf {\bibinfo {volume} {137}},\ \bibinfo {pages}
  {B1628} (\bibinfo {year} {1965})}\BibitemShut {NoStop}%
\bibitem [{\citenamefont {Llewellyn~Smith}(1972)}]{LlewellynSmith:1971uhs}%
  \BibitemOpen
  \bibfield  {author} {\bibinfo {author} {\bibfnamefont {C.~H.}\ \bibnamefont
  {Llewellyn~Smith}},\ }\bibfield  {title} {\bibinfo {title} {{Neutrino
  Reactions at Accelerator Energies}},\ }\href
  {https://doi.org/10.1016/0370-1573(72)90010-5} {\bibfield  {journal}
  {\bibinfo  {journal} {Phys. Rept.}\ }\textbf {\bibinfo {volume} {3}},\
  \bibinfo {pages} {261} (\bibinfo {year} {1972})}\BibitemShut {NoStop}%
\bibitem [{\citenamefont {Cabibbo}\ \emph {et~al.}(2003)\citenamefont
  {Cabibbo}, \citenamefont {Swallow},\ and\ \citenamefont
  {Winston}}]{Cabibbo:2003cu}%
  \BibitemOpen
  \bibfield  {author} {\bibinfo {author} {\bibfnamefont {N.}~\bibnamefont
  {Cabibbo}}, \bibinfo {author} {\bibfnamefont {E.~C.}\ \bibnamefont
  {Swallow}},\ and\ \bibinfo {author} {\bibfnamefont {R.}~\bibnamefont
  {Winston}},\ }\bibfield  {title} {\bibinfo {title} {{Semileptonic hyperon
  decays}},\ }\href {https://doi.org/10.1146/annurev.nucl.53.013103.155258}
  {\bibfield  {journal} {\bibinfo  {journal} {Ann. Rev. Nucl. Part. Sci.}\
  }\textbf {\bibinfo {volume} {53}},\ \bibinfo {pages} {39} (\bibinfo {year}
  {2003})},\ \Eprint {https://arxiv.org/abs/hep-ph/0307298}
  {arXiv:hep-ph/0307298} \BibitemShut {NoStop}%
\bibitem [{\citenamefont {Shanahan}\ \emph {et~al.}(2015)\citenamefont
  {Shanahan}, \citenamefont {Cooke}, \citenamefont {Horsley}, \citenamefont
  {Nakamura}, \citenamefont {Rakow}, \citenamefont {Schierholz}, \citenamefont
  {Thomas}, \citenamefont {Young},\ and\ \citenamefont
  {Zanotti}}]{Shanahan:2015dka}%
  \BibitemOpen
  \bibfield  {author} {\bibinfo {author} {\bibfnamefont {P.~E.}\ \bibnamefont
  {Shanahan}}, \bibinfo {author} {\bibfnamefont {A.~N.}\ \bibnamefont {Cooke}},
  \bibinfo {author} {\bibfnamefont {R.}~\bibnamefont {Horsley}}, \bibinfo
  {author} {\bibfnamefont {Y.}~\bibnamefont {Nakamura}}, \bibinfo {author}
  {\bibfnamefont {P.~E.~L.}\ \bibnamefont {Rakow}}, \bibinfo {author}
  {\bibfnamefont {G.}~\bibnamefont {Schierholz}}, \bibinfo {author}
  {\bibfnamefont {A.~W.}\ \bibnamefont {Thomas}}, \bibinfo {author}
  {\bibfnamefont {R.~D.}\ \bibnamefont {Young}},\ and\ \bibinfo {author}
  {\bibfnamefont {J.~M.}\ \bibnamefont {Zanotti}},\ }\bibfield  {title}
  {\bibinfo {title} {{SU(3) breaking in hyperon transition vector form
  factors}},\ }\href {https://doi.org/10.1103/PhysRevD.92.074029} {\bibfield
  {journal} {\bibinfo  {journal} {Phys. Rev. D}\ }\textbf {\bibinfo {volume}
  {92}},\ \bibinfo {pages} {074029} (\bibinfo {year} {2015})},\ \Eprint
  {https://arxiv.org/abs/1508.06923} {arXiv:1508.06923 [nucl-th]} \BibitemShut
  {NoStop}%
\bibitem [{\citenamefont {Sasaki}(2012)}]{Sasaki:2012ne}%
  \BibitemOpen
  \bibfield  {author} {\bibinfo {author} {\bibfnamefont {S.}~\bibnamefont
  {Sasaki}},\ }\bibfield  {title} {\bibinfo {title} {{Hyperon vector form
  factor from 2+1 flavor lattice QCD}},\ }\href
  {https://doi.org/10.1103/PhysRevD.86.114502} {\bibfield  {journal} {\bibinfo
  {journal} {Phys. Rev. D}\ }\textbf {\bibinfo {volume} {86}},\ \bibinfo
  {pages} {114502} (\bibinfo {year} {2012})},\ \Eprint
  {https://arxiv.org/abs/1209.6115} {arXiv:1209.6115 [hep-lat]} \BibitemShut
  {NoStop}%
\bibitem [{\citenamefont {Sasaki}(2017)}]{Sasaki:2017jue}%
  \BibitemOpen
  \bibfield  {author} {\bibinfo {author} {\bibfnamefont {S.}~\bibnamefont
  {Sasaki}},\ }\bibfield  {title} {\bibinfo {title} {{Continuum limit of
  hyperon vector coupling $f_1(0)$ from 2+1 flavor domain wall QCD}},\ }\href
  {https://doi.org/10.1103/PhysRevD.96.074509} {\bibfield  {journal} {\bibinfo
  {journal} {Phys. Rev. D}\ }\textbf {\bibinfo {volume} {96}},\ \bibinfo
  {pages} {074509} (\bibinfo {year} {2017})},\ \Eprint
  {https://arxiv.org/abs/1708.04008} {arXiv:1708.04008 [hep-lat]} \BibitemShut
  {NoStop}%
\bibitem [{\citenamefont {Zhu}\ \emph {et~al.}(2001)\citenamefont {Zhu},
  \citenamefont {Puglia},\ and\ \citenamefont {Ramsey-Musolf}}]{Zhu:2000zf}%
  \BibitemOpen
  \bibfield  {author} {\bibinfo {author} {\bibfnamefont {S.-L.}\ \bibnamefont
  {Zhu}}, \bibinfo {author} {\bibfnamefont {S.}~\bibnamefont {Puglia}},\ and\
  \bibinfo {author} {\bibfnamefont {M.~J.}\ \bibnamefont {Ramsey-Musolf}},\
  }\bibfield  {title} {\bibinfo {title} {{Recoil order chiral corrections to
  baryon octet axial currents}},\ }\href
  {https://doi.org/10.1103/PhysRevD.63.034002} {\bibfield  {journal} {\bibinfo
  {journal} {Phys. Rev. D}\ }\textbf {\bibinfo {volume} {63}},\ \bibinfo
  {pages} {034002} (\bibinfo {year} {2001})},\ \Eprint
  {https://arxiv.org/abs/hep-ph/0009159} {arXiv:hep-ph/0009159} \BibitemShut
  {NoStop}%
\bibitem [{\citenamefont {Lacour}\ \emph {et~al.}(2007)\citenamefont {Lacour},
  \citenamefont {Kubis},\ and\ \citenamefont {Meissner}}]{Lacour:2007wm}%
  \BibitemOpen
  \bibfield  {author} {\bibinfo {author} {\bibfnamefont {A.}~\bibnamefont
  {Lacour}}, \bibinfo {author} {\bibfnamefont {B.}~\bibnamefont {Kubis}},\ and\
  \bibinfo {author} {\bibfnamefont {U.-G.}\ \bibnamefont {Meissner}},\
  }\bibfield  {title} {\bibinfo {title} {{Hyperon decay form-factors in chiral
  perturbation theory}},\ }\href
  {https://doi.org/10.1088/1126-6708/2007/10/083} {\bibfield  {journal}
  {\bibinfo  {journal} {JHEP}\ }\textbf {\bibinfo {volume} {10}},\ \bibinfo
  {pages} {083}},\ \Eprint {https://arxiv.org/abs/0708.3957} {arXiv:0708.3957
  [hep-ph]} \BibitemShut {NoStop}%
\bibitem [{\citenamefont {Ledwig}\ \emph {et~al.}(2014)\citenamefont {Ledwig},
  \citenamefont {Martin~Camalich}, \citenamefont {Geng},\ and\ \citenamefont
  {Vicente~Vacas}}]{Ledwig:2014rfa}%
  \BibitemOpen
  \bibfield  {author} {\bibinfo {author} {\bibfnamefont {T.}~\bibnamefont
  {Ledwig}}, \bibinfo {author} {\bibfnamefont {J.}~\bibnamefont
  {Martin~Camalich}}, \bibinfo {author} {\bibfnamefont {L.~S.}\ \bibnamefont
  {Geng}},\ and\ \bibinfo {author} {\bibfnamefont {M.~J.}\ \bibnamefont
  {Vicente~Vacas}},\ }\bibfield  {title} {\bibinfo {title} {{Octet-baryon
  axial-vector charges and SU(3)-breaking effects in the semileptonic hyperon
  decays}},\ }\href {https://doi.org/10.1103/PhysRevD.90.054502} {\bibfield
  {journal} {\bibinfo  {journal} {Phys. Rev. D}\ }\textbf {\bibinfo {volume}
  {90}},\ \bibinfo {pages} {054502} (\bibinfo {year} {2014})},\ \Eprint
  {https://arxiv.org/abs/1405.5456} {arXiv:1405.5456 [hep-ph]} \BibitemShut
  {NoStop}%
\bibitem [{\citenamefont {Sauerwein}\ \emph {et~al.}(2022)\citenamefont
  {Sauerwein}, \citenamefont {Lutz},\ and\ \citenamefont
  {Timmermans}}]{Sauerwein:2021jxb}%
  \BibitemOpen
  \bibfield  {author} {\bibinfo {author} {\bibfnamefont {U.}~\bibnamefont
  {Sauerwein}}, \bibinfo {author} {\bibfnamefont {M.~F.~M.}\ \bibnamefont
  {Lutz}},\ and\ \bibinfo {author} {\bibfnamefont {R.~G.~E.}\ \bibnamefont
  {Timmermans}},\ }\bibfield  {title} {\bibinfo {title} {{Axial-vector form
  factors of the baryon octet and chiral symmetry}},\ }\href
  {https://doi.org/10.1103/PhysRevD.105.054005} {\bibfield  {journal} {\bibinfo
   {journal} {Phys. Rev. D}\ }\textbf {\bibinfo {volume} {105}},\ \bibinfo
  {pages} {054005} (\bibinfo {year} {2022})},\ \Eprint
  {https://arxiv.org/abs/2105.06755} {arXiv:2105.06755 [hep-ph]} \BibitemShut
  {NoStop}%
\bibitem [{\citenamefont {Flores-Mendieta}\ \emph {et~al.}(1998)\citenamefont
  {Flores-Mendieta}, \citenamefont {Jenkins},\ and\ \citenamefont
  {Manohar}}]{Flores-Mendieta:1998tfv}%
  \BibitemOpen
  \bibfield  {author} {\bibinfo {author} {\bibfnamefont {R.}~\bibnamefont
  {Flores-Mendieta}}, \bibinfo {author} {\bibfnamefont {E.~E.}\ \bibnamefont
  {Jenkins}},\ and\ \bibinfo {author} {\bibfnamefont {A.~V.}\ \bibnamefont
  {Manohar}},\ }\bibfield  {title} {\bibinfo {title} {{SU(3) symmetry breaking
  in hyperon semileptonic decays}},\ }\href
  {https://doi.org/10.1103/PhysRevD.58.094028} {\bibfield  {journal} {\bibinfo
  {journal} {Phys. Rev. D}\ }\textbf {\bibinfo {volume} {58}},\ \bibinfo
  {pages} {094028} (\bibinfo {year} {1998})},\ \Eprint
  {https://arxiv.org/abs/hep-ph/9805416} {arXiv:hep-ph/9805416} \BibitemShut
  {NoStop}%
\bibitem [{\citenamefont {Buchmann}\ and\ \citenamefont
  {Lebed}(2003)}]{Buchmann:2002et}%
  \BibitemOpen
  \bibfield  {author} {\bibinfo {author} {\bibfnamefont {A.~J.}\ \bibnamefont
  {Buchmann}}\ and\ \bibinfo {author} {\bibfnamefont {R.~F.}\ \bibnamefont
  {Lebed}},\ }\bibfield  {title} {\bibinfo {title} {{Baryon charge radii and
  quadrupole moments in the 1/N(c) expansion: The three flavor case}},\ }\href
  {https://doi.org/10.1103/PhysRevD.67.016002} {\bibfield  {journal} {\bibinfo
  {journal} {Phys. Rev. D}\ }\textbf {\bibinfo {volume} {67}},\ \bibinfo
  {pages} {016002} (\bibinfo {year} {2003})},\ \Eprint
  {https://arxiv.org/abs/hep-ph/0207358} {arXiv:hep-ph/0207358} \BibitemShut
  {NoStop}%
\bibitem [{\citenamefont {Calle~Cordon}\ and\ \citenamefont
  {Goity}(2013)}]{CalleCordon:2012xz}%
  \BibitemOpen
  \bibfield  {author} {\bibinfo {author} {\bibfnamefont {A.}~\bibnamefont
  {Calle~Cordon}}\ and\ \bibinfo {author} {\bibfnamefont {J.~L.}\ \bibnamefont
  {Goity}},\ }\bibfield  {title} {\bibinfo {title} {{Baryon Masses and Axial
  Couplings in the Combined 1/$N_c$ and Chiral Expansions}},\ }\href
  {https://doi.org/10.1103/PhysRevD.87.016019} {\bibfield  {journal} {\bibinfo
  {journal} {Phys. Rev. D}\ }\textbf {\bibinfo {volume} {87}},\ \bibinfo
  {pages} {016019} (\bibinfo {year} {2013})},\ \Eprint
  {https://arxiv.org/abs/1210.2364} {arXiv:1210.2364 [nucl-th]} \BibitemShut
  {NoStop}%
\bibitem [{\citenamefont {Schlumpf}(1995)}]{Schlumpf:1994fb}%
  \BibitemOpen
  \bibfield  {author} {\bibinfo {author} {\bibfnamefont {F.}~\bibnamefont
  {Schlumpf}},\ }\bibfield  {title} {\bibinfo {title} {{Beta decay of hyperons
  in a relativistic quark model}},\ }\href
  {https://doi.org/10.1103/PhysRevD.51.2262} {\bibfield  {journal} {\bibinfo
  {journal} {Phys. Rev. D}\ }\textbf {\bibinfo {volume} {51}},\ \bibinfo
  {pages} {2262} (\bibinfo {year} {1995})},\ \Eprint
  {https://arxiv.org/abs/hep-ph/9409272} {arXiv:hep-ph/9409272} \BibitemShut
  {NoStop}%
\bibitem [{\citenamefont {Ramalho}\ and\ \citenamefont
  {Tsushima}(2016)}]{Ramalho:2015jem}%
  \BibitemOpen
  \bibfield  {author} {\bibinfo {author} {\bibfnamefont {G.}~\bibnamefont
  {Ramalho}}\ and\ \bibinfo {author} {\bibfnamefont {K.}~\bibnamefont
  {Tsushima}},\ }\bibfield  {title} {\bibinfo {title} {{Axial form factors of
  the octet baryons in a covariant quark model}},\ }\href
  {https://doi.org/10.1103/PhysRevD.94.014001} {\bibfield  {journal} {\bibinfo
  {journal} {Phys. Rev. D}\ }\textbf {\bibinfo {volume} {94}},\ \bibinfo
  {pages} {014001} (\bibinfo {year} {2016})},\ \Eprint
  {https://arxiv.org/abs/1512.01167} {arXiv:1512.01167 [hep-ph]} \BibitemShut
  {NoStop}%
\bibitem [{\citenamefont {Yang}\ and\ \citenamefont
  {Kim}(2015)}]{Yang:2015era}%
  \BibitemOpen
  \bibfield  {author} {\bibinfo {author} {\bibfnamefont {G.-S.}\ \bibnamefont
  {Yang}}\ and\ \bibinfo {author} {\bibfnamefont {H.-C.}\ \bibnamefont {Kim}},\
  }\bibfield  {title} {\bibinfo {title} {{Hyperon Semileptonic decay constants
  with flavor SU(3) symmetry breaking}},\ }\href
  {https://doi.org/10.1103/PhysRevC.92.035206} {\bibfield  {journal} {\bibinfo
  {journal} {Phys. Rev. C}\ }\textbf {\bibinfo {volume} {92}},\ \bibinfo
  {pages} {035206} (\bibinfo {year} {2015})},\ \Eprint
  {https://arxiv.org/abs/1504.04453} {arXiv:1504.04453 [hep-ph]} \BibitemShut
  {NoStop}%
\bibitem [{\citenamefont {Liu}\ \emph {et~al.}(2023)\citenamefont {Liu},
  \citenamefont {Limphirat}, \citenamefont {Xu}, \citenamefont {Zhao},
  \citenamefont {Khosonthongkee},\ and\ \citenamefont {Yan}}]{Liu:2022ekr}%
  \BibitemOpen
  \bibfield  {author} {\bibinfo {author} {\bibfnamefont {X.~Y.}\ \bibnamefont
  {Liu}}, \bibinfo {author} {\bibfnamefont {A.}~\bibnamefont {Limphirat}},
  \bibinfo {author} {\bibfnamefont {K.}~\bibnamefont {Xu}}, \bibinfo {author}
  {\bibfnamefont {Z.}~\bibnamefont {Zhao}}, \bibinfo {author} {\bibfnamefont
  {K.}~\bibnamefont {Khosonthongkee}},\ and\ \bibinfo {author} {\bibfnamefont
  {Y.}~\bibnamefont {Yan}},\ }\bibfield  {title} {\bibinfo {title} {{Axial
  transition form factors of octet baryons in the perturbative chiral quark
  model}},\ }\href {https://doi.org/10.1103/PhysRevD.107.074006} {\bibfield
  {journal} {\bibinfo  {journal} {Phys. Rev. D}\ }\textbf {\bibinfo {volume}
  {107}},\ \bibinfo {pages} {074006} (\bibinfo {year} {2023})},\ \Eprint
  {https://arxiv.org/abs/2209.00808} {arXiv:2209.00808 [hep-ph]} \BibitemShut
  {NoStop}%
\bibitem [{\citenamefont {Eichten}\ \emph {et~al.}(1972)\citenamefont {Eichten}
  \emph {et~al.}}]{Eichten:1972bb}%
  \BibitemOpen
  \bibfield  {author} {\bibinfo {author} {\bibfnamefont {T.}~\bibnamefont
  {Eichten}} \emph {et~al.},\ }\bibfield  {title} {\bibinfo {title}
  {{Observation of 'Elastic' Hyperon Production by Anti-neutrinos}},\ }\href
  {https://doi.org/10.1016/0370-2693(72)90490-X} {\bibfield  {journal}
  {\bibinfo  {journal} {Phys. Lett. B}\ }\textbf {\bibinfo {volume} {40}},\
  \bibinfo {pages} {593} (\bibinfo {year} {1972})}\BibitemShut {NoStop}%
\bibitem [{\citenamefont {Erriquez}\ \emph {et~al.}(1977)\citenamefont
  {Erriquez} \emph {et~al.}}]{Erriquez:1977tr}%
  \BibitemOpen
  \bibfield  {author} {\bibinfo {author} {\bibfnamefont {O.}~\bibnamefont
  {Erriquez}} \emph {et~al.},\ }\bibfield  {title} {\bibinfo {title} {{Strange
  Particle Production by anti-neutrinos}},\ }\href
  {https://doi.org/10.1016/0370-2693(77)90683-9} {\bibfield  {journal}
  {\bibinfo  {journal} {Phys. Lett. B}\ }\textbf {\bibinfo {volume} {70}},\
  \bibinfo {pages} {383} (\bibinfo {year} {1977})}\BibitemShut {NoStop}%
\bibitem [{\citenamefont {Erriquez}\ \emph {et~al.}(1978)\citenamefont
  {Erriquez} \emph {et~al.}}]{Erriquez:1978pg}%
  \BibitemOpen
  \bibfield  {author} {\bibinfo {author} {\bibfnamefont {O.}~\bibnamefont
  {Erriquez}} \emph {et~al.},\ }\bibfield  {title} {\bibinfo {title}
  {{Production of Strange Particles in anti-neutrino Interactions at the CERN
  PS}},\ }\href {https://doi.org/10.1016/0550-3213(78)90316-4} {\bibfield
  {journal} {\bibinfo  {journal} {Nucl. Phys. B}\ }\textbf {\bibinfo {volume}
  {140}},\ \bibinfo {pages} {123} (\bibinfo {year} {1978})}\BibitemShut
  {NoStop}%
\bibitem [{\citenamefont {Barish}\ \emph {et~al.}(1974)\citenamefont {Barish}
  \emph {et~al.}}]{Barish:1974ye}%
  \BibitemOpen
  \bibfield  {author} {\bibinfo {author} {\bibfnamefont {S.~J.}\ \bibnamefont
  {Barish}} \emph {et~al.},\ }\bibfield  {title} {\bibinfo {title}
  {{Strange-Particle Production in Neutrino Interactions}},\ }\href
  {https://doi.org/10.1103/PhysRevLett.33.1446} {\bibfield  {journal} {\bibinfo
   {journal} {Phys. Rev. Lett.}\ }\textbf {\bibinfo {volume} {33}},\ \bibinfo
  {pages} {1446} (\bibinfo {year} {1974})}\BibitemShut {NoStop}%
\bibitem [{\citenamefont {Fanourakis}\ \emph {et~al.}(1980)\citenamefont
  {Fanourakis}, \citenamefont {Resvanis}, \citenamefont {Grammatikakis},
  \citenamefont {Tsilimigras}, \citenamefont {Vayaki}, \citenamefont
  {Camerini}, \citenamefont {Fry}, \citenamefont {Loveless}, \citenamefont
  {Mapp},\ and\ \citenamefont {Reeder}}]{Fanourakis:1980si}%
  \BibitemOpen
  \bibfield  {author} {\bibinfo {author} {\bibfnamefont {G.}~\bibnamefont
  {Fanourakis}}, \bibinfo {author} {\bibfnamefont {L.~K.}\ \bibnamefont
  {Resvanis}}, \bibinfo {author} {\bibfnamefont {G.}~\bibnamefont
  {Grammatikakis}}, \bibinfo {author} {\bibfnamefont {P.}~\bibnamefont
  {Tsilimigras}}, \bibinfo {author} {\bibfnamefont {A.}~\bibnamefont {Vayaki}},
  \bibinfo {author} {\bibfnamefont {U.}~\bibnamefont {Camerini}}, \bibinfo
  {author} {\bibfnamefont {W.~F.}\ \bibnamefont {Fry}}, \bibinfo {author}
  {\bibfnamefont {R.~J.}\ \bibnamefont {Loveless}}, \bibinfo {author}
  {\bibfnamefont {J.~H.}\ \bibnamefont {Mapp}},\ and\ \bibinfo {author}
  {\bibfnamefont {D.~D.}\ \bibnamefont {Reeder}},\ }\bibfield  {title}
  {\bibinfo {title} {{Study of Low-energy Anti-neutrino Interactions on
  Protons}},\ }\href {https://doi.org/10.1103/PhysRevD.21.562} {\bibfield
  {journal} {\bibinfo  {journal} {Phys. Rev. D}\ }\textbf {\bibinfo {volume}
  {21}},\ \bibinfo {pages} {562} (\bibinfo {year} {1980})}\BibitemShut
  {NoStop}%
\bibitem [{\citenamefont {Baker}\ \emph {et~al.}(1981)\citenamefont {Baker},
  \citenamefont {Connolly}, \citenamefont {Kahn}, \citenamefont {Kirk},
  \citenamefont {Murtagh}, \citenamefont {Palmer}, \citenamefont {Samios},\
  and\ \citenamefont {Tanaka}}]{Baker:1981tx}%
  \BibitemOpen
  \bibfield  {author} {\bibinfo {author} {\bibfnamefont {N.~J.}\ \bibnamefont
  {Baker}}, \bibinfo {author} {\bibfnamefont {P.~L.}\ \bibnamefont {Connolly}},
  \bibinfo {author} {\bibfnamefont {S.~A.}\ \bibnamefont {Kahn}}, \bibinfo
  {author} {\bibfnamefont {H.~G.}\ \bibnamefont {Kirk}}, \bibinfo {author}
  {\bibfnamefont {M.~J.}\ \bibnamefont {Murtagh}}, \bibinfo {author}
  {\bibfnamefont {R.~B.}\ \bibnamefont {Palmer}}, \bibinfo {author}
  {\bibfnamefont {N.~P.}\ \bibnamefont {Samios}},\ and\ \bibinfo {author}
  {\bibfnamefont {M.}~\bibnamefont {Tanaka}},\ }\bibfield  {title} {\bibinfo
  {title} {{Strange Particle Production from Neutrino Interactions in the BNL
  7-Ft Bubble Chamber}},\ }\href {https://doi.org/10.1103/PhysRevD.24.2779}
  {\bibfield  {journal} {\bibinfo  {journal} {Phys. Rev. D}\ }\textbf {\bibinfo
  {volume} {24}},\ \bibinfo {pages} {2779} (\bibinfo {year}
  {1981})}\BibitemShut {NoStop}%
\bibitem [{\citenamefont {Ammosov}\ \emph {et~al.}(1987)\citenamefont {Ammosov}
  \emph {et~al.}}]{Ammosov:1986jn}%
  \BibitemOpen
  \bibfield  {author} {\bibinfo {author} {\bibfnamefont {V.~V.}\ \bibnamefont
  {Ammosov}} \emph {et~al.},\ }\bibfield  {title} {\bibinfo {title} {{Neutral
  Strange Particle Exclusive Production in Charged Current High-energy
  Anti-neutrino Interactions}},\ }\href {https://doi.org/10.1007/BF01573931}
  {\bibfield  {journal} {\bibinfo  {journal} {Z. Phys. C}\ }\textbf {\bibinfo
  {volume} {36}},\ \bibinfo {pages} {377} (\bibinfo {year} {1987})}\BibitemShut
  {NoStop}%
\bibitem [{\citenamefont {Son}\ \emph {et~al.}(1983)\citenamefont {Son} \emph
  {et~al.}}]{Son:1983xh}%
  \BibitemOpen
  \bibfield  {author} {\bibinfo {author} {\bibfnamefont {D.}~\bibnamefont
  {Son}} \emph {et~al.},\ }\bibfield  {title} {\bibinfo {title} {{Quasielastic
  Charmed Baryon Production and Exclusive Strange Particle Production by
  High-energy Neutrinos}},\ }\href {https://doi.org/10.1103/PhysRevD.28.2129}
  {\bibfield  {journal} {\bibinfo  {journal} {Phys. Rev. D}\ }\textbf {\bibinfo
  {volume} {28}},\ \bibinfo {pages} {2129} (\bibinfo {year}
  {1983})}\BibitemShut {NoStop}%
\bibitem [{\citenamefont {Brunner}\ \emph {et~al.}(1990)\citenamefont {Brunner}
  \emph {et~al.}}]{SKAT:1989nel}%
  \BibitemOpen
  \bibfield  {author} {\bibinfo {author} {\bibfnamefont {J.}~\bibnamefont
  {Brunner}} \emph {et~al.} (\bibinfo {collaboration} {SKAT}),\ }\bibfield
  {title} {\bibinfo {title} {{Quasielastic Nucleon and Hyperon Production by
  Neutrinos and Anti-neutrinos With Energies Below 30-{GeV}}},\ }\href
  {https://doi.org/10.1007/BF01556267} {\bibfield  {journal} {\bibinfo
  {journal} {Z. Phys. C}\ }\textbf {\bibinfo {volume} {45}},\ \bibinfo {pages}
  {551} (\bibinfo {year} {1990})}\BibitemShut {NoStop}%
\bibitem [{\citenamefont {Formaggio}\ and\ \citenamefont
  {Zeller}(2012)}]{Formaggio:2012cpf}%
  \BibitemOpen
  \bibfield  {author} {\bibinfo {author} {\bibfnamefont {J.~A.}\ \bibnamefont
  {Formaggio}}\ and\ \bibinfo {author} {\bibfnamefont {G.~P.}\ \bibnamefont
  {Zeller}},\ }\bibfield  {title} {\bibinfo {title} {{From eV to EeV: Neutrino
  Cross Sections Across Energy Scales}},\ }\href
  {https://doi.org/10.1103/RevModPhys.84.1307} {\bibfield  {journal} {\bibinfo
  {journal} {Rev. Mod. Phys.}\ }\textbf {\bibinfo {volume} {84}},\ \bibinfo
  {pages} {1307} (\bibinfo {year} {2012})},\ \Eprint
  {https://arxiv.org/abs/1305.7513} {arXiv:1305.7513 [hep-ex]} \BibitemShut
  {NoStop}%
\bibitem [{\citenamefont {Astier}\ \emph {et~al.}(2000)\citenamefont {Astier}
  \emph {et~al.}}]{NOMAD:2000wdf}%
  \BibitemOpen
  \bibfield  {author} {\bibinfo {author} {\bibfnamefont {P.}~\bibnamefont
  {Astier}} \emph {et~al.} (\bibinfo {collaboration} {NOMAD}),\ }\bibfield
  {title} {\bibinfo {title} {{Measurement of the Lambda polarization in nu/mu
  charged current interactions in the NOMAD experiment}},\ }\href
  {https://doi.org/10.1016/S0550-3213(00)00503-4} {\bibfield  {journal}
  {\bibinfo  {journal} {Nucl. Phys. B}\ }\textbf {\bibinfo {volume} {588}},\
  \bibinfo {pages} {3} (\bibinfo {year} {2000})}\BibitemShut {NoStop}%
\bibitem [{\citenamefont {Naumov}\ \emph {et~al.}(2004)\citenamefont {Naumov}
  \emph {et~al.}}]{NOMAD:2004djf}%
  \BibitemOpen
  \bibfield  {author} {\bibinfo {author} {\bibfnamefont {D.}~\bibnamefont
  {Naumov}} \emph {et~al.} (\bibinfo {collaboration} {NOMAD}),\ }\bibfield
  {title} {\bibinfo {title} {{A Study of strange particles produced in neutrino
  neutral current interactions in the NOMAD experiment}},\ }\href
  {https://doi.org/10.1016/j.nuclphysb.2004.09.013} {\bibfield  {journal}
  {\bibinfo  {journal} {Nucl. Phys. B}\ }\textbf {\bibinfo {volume} {700}},\
  \bibinfo {pages} {51} (\bibinfo {year} {2004})},\ \Eprint
  {https://arxiv.org/abs/hep-ex/0409037} {arXiv:hep-ex/0409037} \BibitemShut
  {NoStop}%
\bibitem [{\citenamefont {Abratenko}\ \emph {et~al.}(2022)\citenamefont
  {Abratenko} \emph {et~al.}}]{MicroBooNE:2022bpw}%
  \BibitemOpen
  \bibfield  {author} {\bibinfo {author} {\bibfnamefont {P.}~\bibnamefont
  {Abratenko}} \emph {et~al.} (\bibinfo {collaboration} {MicroBooNE}),\
  }\bibfield  {title} {\bibinfo {title} {{First measurement of quasi-elastic
  $\Lambda$ baryon production in muon anti-neutrino interactions in the
  MicroBooNE detector}},\ }\href@noop {} {\  (\bibinfo {year} {2022})},\
  \Eprint {https://arxiv.org/abs/2212.07888} {arXiv:2212.07888 [hep-ex]}
  \BibitemShut {NoStop}%
\bibitem [{\citenamefont {Brailsford}(2017)}]{Brailsford:2017rxe}%
  \BibitemOpen
  \bibfield  {author} {\bibinfo {author} {\bibfnamefont {D.}~\bibnamefont
  {Brailsford}} (\bibinfo {collaboration} {SBND}),\ }\bibfield  {title}
  {\bibinfo {title} {{Physics Program of the Short-Baseline Near Detector}},\
  }\href {https://doi.org/10.1088/1742-6596/888/1/012186} {\bibfield  {journal}
  {\bibinfo  {journal} {J. Phys. Conf. Ser.}\ }\textbf {\bibinfo {volume}
  {888}},\ \bibinfo {pages} {012186} (\bibinfo {year} {2017})}\BibitemShut
  {NoStop}%
\bibitem [{\citenamefont {Machado}\ \emph {et~al.}(2019)\citenamefont
  {Machado}, \citenamefont {Palamara},\ and\ \citenamefont
  {Schmitz}}]{Machado:2019oxb}%
  \BibitemOpen
  \bibfield  {author} {\bibinfo {author} {\bibfnamefont {P.~A.}\ \bibnamefont
  {Machado}}, \bibinfo {author} {\bibfnamefont {O.}~\bibnamefont {Palamara}},\
  and\ \bibinfo {author} {\bibfnamefont {D.~W.}\ \bibnamefont {Schmitz}},\
  }\bibfield  {title} {\bibinfo {title} {{The Short-Baseline Neutrino Program
  at Fermilab}},\ }\href {https://doi.org/10.1146/annurev-nucl-101917-020949}
  {\bibfield  {journal} {\bibinfo  {journal} {Ann. Rev. Nucl. Part. Sci.}\
  }\textbf {\bibinfo {volume} {69}},\ \bibinfo {pages} {363} (\bibinfo {year}
  {2019})},\ \Eprint {https://arxiv.org/abs/1903.04608} {arXiv:1903.04608
  [hep-ex]} \BibitemShut {NoStop}%
\bibitem [{\citenamefont {Singh}\ and\ \citenamefont
  {Vicente~Vacas}(2006)}]{Singh:2006xp}%
  \BibitemOpen
  \bibfield  {author} {\bibinfo {author} {\bibfnamefont {S.~K.}\ \bibnamefont
  {Singh}}\ and\ \bibinfo {author} {\bibfnamefont {M.~J.}\ \bibnamefont
  {Vicente~Vacas}},\ }\bibfield  {title} {\bibinfo {title} {{Weak quasi-elastic
  production of hyperons}},\ }\href
  {https://doi.org/10.1103/PhysRevD.74.053009} {\bibfield  {journal} {\bibinfo
  {journal} {Phys.Rev.}\ }\textbf {\bibinfo {volume} {D74}},\ \bibinfo {pages}
  {053009} (\bibinfo {year} {2006})}\BibitemShut {NoStop}%
\bibitem [{\citenamefont {Rafi~Alam}\ \emph {et~al.}(2013)\citenamefont
  {Rafi~Alam}, \citenamefont {Chauhan}, \citenamefont {Sajjad~Athar},\ and\
  \citenamefont {Singh}}]{RafiAlam:2013prd}%
  \BibitemOpen
  \bibfield  {author} {\bibinfo {author} {\bibfnamefont {M.}~\bibnamefont
  {Rafi~Alam}}, \bibinfo {author} {\bibfnamefont {S.}~\bibnamefont {Chauhan}},
  \bibinfo {author} {\bibfnamefont {M.}~\bibnamefont {Sajjad~Athar}},\ and\
  \bibinfo {author} {\bibfnamefont {S.~K.}\ \bibnamefont {Singh}},\ }\bibfield
  {title} {\bibinfo {title} {{$\bar{\nu}_l$ induced pion production from nuclei
  at $\sim{1}$ GeV}},\ }\href {https://doi.org/10.1103/PhysRevD.88.077301}
  {\bibfield  {journal} {\bibinfo  {journal} {Phys. Rev. D}\ }\textbf {\bibinfo
  {volume} {88}},\ \bibinfo {pages} {077301} (\bibinfo {year} {2013})},\
  \Eprint {https://arxiv.org/abs/1310.7704} {arXiv:1310.7704 [nucl-th]}
  \BibitemShut {NoStop}%
\bibitem [{\citenamefont {Fatima}\ \emph {et~al.}(2021)\citenamefont {Fatima},
  \citenamefont {Athar},\ and\ \citenamefont {Singh}}]{Fatima:2021ctt}%
  \BibitemOpen
  \bibfield  {author} {\bibinfo {author} {\bibfnamefont {A.}~\bibnamefont
  {Fatima}}, \bibinfo {author} {\bibfnamefont {M.~S.}\ \bibnamefont {Athar}},\
  and\ \bibinfo {author} {\bibfnamefont {S.~K.}\ \bibnamefont {Singh}},\
  }\bibfield  {title} {\bibinfo {title} {{${\bar{\nu }}_{\mu }$ induced
  quasielastic production of hyperons leading to pions}},\ }\href
  {https://doi.org/10.1140/epjs/s11734-021-00272-0} {\bibfield  {journal}
  {\bibinfo  {journal} {Eur. Phys. J. ST}\ }\textbf {\bibinfo {volume} {230}},\
  \bibinfo {pages} {4391} (\bibinfo {year} {2021})},\ \Eprint
  {https://arxiv.org/abs/2106.14590} {arXiv:2106.14590 [hep-ph]} \BibitemShut
  {NoStop}%
\bibitem [{\citenamefont {Mintz}\ and\ \citenamefont
  {Wen}(2006)}]{Mintz:2006yp}%
  \BibitemOpen
  \bibfield  {author} {\bibinfo {author} {\bibfnamefont {S.~L.}\ \bibnamefont
  {Mintz}}\ and\ \bibinfo {author} {\bibfnamefont {L.-L.}\ \bibnamefont
  {Wen}},\ }\bibfield  {title} {\bibinfo {title} {{The weak production of
  Lambda particles in antineutrino-proton scattering and the contributions of
  the form factors}},\ }\href {https://doi.org/10.1016/j.nuclphysa.2005.11.024}
  {\bibfield  {journal} {\bibinfo  {journal} {Nucl. Phys. A}\ }\textbf
  {\bibinfo {volume} {766}},\ \bibinfo {pages} {219} (\bibinfo {year}
  {2006})}\BibitemShut {NoStop}%
\bibitem [{\citenamefont {Kuzmin}\ and\ \citenamefont
  {Naumov}(2009)}]{Kuzmin:2008zz}%
  \BibitemOpen
  \bibfield  {author} {\bibinfo {author} {\bibfnamefont {K.~S.}\ \bibnamefont
  {Kuzmin}}\ and\ \bibinfo {author} {\bibfnamefont {V.~A.}\ \bibnamefont
  {Naumov}},\ }\bibfield  {title} {\bibinfo {title} {{Axial mass in reactions
  of quasielastic antineutrino-nucleon scattering with strange hyperon
  production}},\ }\href {https://doi.org/10.1134/S1063778809090105} {\bibfield
  {journal} {\bibinfo  {journal} {Phys. Atom. Nucl.}\ }\textbf {\bibinfo
  {volume} {72}},\ \bibinfo {pages} {1501} (\bibinfo {year}
  {2009})}\BibitemShut {NoStop}%
\bibitem [{\citenamefont {Dewan}(1981)}]{Dewan:1981ab}%
  \BibitemOpen
  \bibfield  {author} {\bibinfo {author} {\bibfnamefont {H.~K.}\ \bibnamefont
  {Dewan}},\ }\bibfield  {title} {\bibinfo {title} {{Strange Particle
  Production in Neutrino Scattering}},\ }\href
  {https://doi.org/10.1103/PhysRevD.24.2369} {\bibfield  {journal} {\bibinfo
  {journal} {Phys. Rev. D}\ }\textbf {\bibinfo {volume} {24}},\ \bibinfo
  {pages} {2369} (\bibinfo {year} {1981})}\BibitemShut {NoStop}%
\bibitem [{\citenamefont {Benitez~Galan}\ \emph {et~al.}(2021)\citenamefont
  {Benitez~Galan}, \citenamefont {Alam},\ and\ \citenamefont
  {Ruiz~Simo}}]{BenitezGalan:2021jdm}%
  \BibitemOpen
  \bibfield  {author} {\bibinfo {author} {\bibfnamefont {M.}~\bibnamefont
  {Benitez~Galan}}, \bibinfo {author} {\bibfnamefont {M.~R.}\ \bibnamefont
  {Alam}},\ and\ \bibinfo {author} {\bibfnamefont {I.}~\bibnamefont
  {Ruiz~Simo}},\ }\bibfield  {title} {\bibinfo {title} {{Cabibbo suppressed
  single pion production off the nucleon induced by antineutrinos}},\ }\href
  {https://doi.org/10.1103/PhysRevD.104.073005} {\bibfield  {journal} {\bibinfo
   {journal} {Phys. Rev. D}\ }\textbf {\bibinfo {volume} {104}},\ \bibinfo
  {pages} {073005} (\bibinfo {year} {2021})},\ \Eprint
  {https://arxiv.org/abs/2108.06393} {arXiv:2108.06393 [hep-ph]} \BibitemShut
  {NoStop}%
\bibitem [{\citenamefont {Ren}\ \emph {et~al.}(2015)\citenamefont {Ren},
  \citenamefont {Oset}, \citenamefont {Alvarez-Ruso},\ and\ \citenamefont
  {Vicente~Vacas}}]{Ren:2015bsa}%
  \BibitemOpen
  \bibfield  {author} {\bibinfo {author} {\bibfnamefont {X.-L.}\ \bibnamefont
  {Ren}}, \bibinfo {author} {\bibfnamefont {E.}~\bibnamefont {Oset}}, \bibinfo
  {author} {\bibfnamefont {L.}~\bibnamefont {Alvarez-Ruso}},\ and\ \bibinfo
  {author} {\bibfnamefont {M.~J.}\ \bibnamefont {Vicente~Vacas}},\ }\bibfield
  {title} {\bibinfo {title} {{Antineutrino induced \ensuremath{\Lambda}(1405)
  production off the proton}},\ }\href
  {https://doi.org/10.1103/PhysRevC.91.045201} {\bibfield  {journal} {\bibinfo
  {journal} {Phys. Rev. C}\ }\textbf {\bibinfo {volume} {91}},\ \bibinfo
  {pages} {045201} (\bibinfo {year} {2015})},\ \Eprint
  {https://arxiv.org/abs/1501.04073} {arXiv:1501.04073 [hep-ph]} \BibitemShut
  {NoStop}%
\bibitem [{\citenamefont {Alam}\ \emph {et~al.}(2012)\citenamefont {Alam},
  \citenamefont {Simo}, \citenamefont {Athar},\ and\ \citenamefont
  {Vicente~Vacas}}]{Alam:2011vwg}%
  \BibitemOpen
  \bibfield  {author} {\bibinfo {author} {\bibfnamefont {M.~R.}\ \bibnamefont
  {Alam}}, \bibinfo {author} {\bibfnamefont {I.~R.}\ \bibnamefont {Simo}},
  \bibinfo {author} {\bibfnamefont {M.~S.}\ \bibnamefont {Athar}},\ and\
  \bibinfo {author} {\bibfnamefont {M.~J.}\ \bibnamefont {Vicente~Vacas}},\
  }\bibfield  {title} {\bibinfo {title} {{$\bar{\nu}$ induced $\bar{K}$
  production off the nucleon}},\ }\href
  {https://doi.org/10.1103/PhysRevD.85.013014} {\bibfield  {journal} {\bibinfo
  {journal} {Phys. Rev. D}\ }\textbf {\bibinfo {volume} {85}},\ \bibinfo
  {pages} {013014} (\bibinfo {year} {2012})},\ \Eprint
  {https://arxiv.org/abs/1111.0863} {arXiv:1111.0863 [hep-ph]} \BibitemShut
  {NoStop}%
\bibitem [{\citenamefont {Shrock}(1975)}]{Shrock:1975an}%
  \BibitemOpen
  \bibfield  {author} {\bibinfo {author} {\bibfnamefont {R.~E.}\ \bibnamefont
  {Shrock}},\ }\bibfield  {title} {\bibinfo {title} {{Associated Production by
  Weak Charged and Neutral Currents}},\ }\href
  {https://doi.org/10.1103/PhysRevD.12.2049} {\bibfield  {journal} {\bibinfo
  {journal} {Phys. Rev. D}\ }\textbf {\bibinfo {volume} {12}},\ \bibinfo
  {pages} {2049} (\bibinfo {year} {1975})}\BibitemShut {NoStop}%
\bibitem [{\citenamefont {Adera}\ \emph {et~al.}(2010)\citenamefont {Adera},
  \citenamefont {Van Der~Ventel}, \citenamefont {van Niekerk},\ and\
  \citenamefont {Mart}}]{Adera:2010zz}%
  \BibitemOpen
  \bibfield  {author} {\bibinfo {author} {\bibfnamefont {G.~B.}\ \bibnamefont
  {Adera}}, \bibinfo {author} {\bibfnamefont {B.~I.~S.}\ \bibnamefont {Van
  Der~Ventel}}, \bibinfo {author} {\bibfnamefont {D.~D.}\ \bibnamefont {van
  Niekerk}},\ and\ \bibinfo {author} {\bibfnamefont {T.}~\bibnamefont {Mart}},\
  }\bibfield  {title} {\bibinfo {title} {{Strange-particle production via the
  weak interaction}},\ }\href {https://doi.org/10.1103/PhysRevC.82.025501}
  {\bibfield  {journal} {\bibinfo  {journal} {Phys. Rev. C}\ }\textbf {\bibinfo
  {volume} {82}},\ \bibinfo {pages} {025501} (\bibinfo {year} {2010})},\
  \Eprint {https://arxiv.org/abs/1112.5748} {arXiv:1112.5748 [nucl-th]}
  \BibitemShut {NoStop}%
\bibitem [{\citenamefont {Nakamura}\ \emph {et~al.}(2015)\citenamefont
  {Nakamura}, \citenamefont {Kamano},\ and\ \citenamefont
  {Sato}}]{Nakamura:2015rta}%
  \BibitemOpen
  \bibfield  {author} {\bibinfo {author} {\bibfnamefont {S.~X.}\ \bibnamefont
  {Nakamura}}, \bibinfo {author} {\bibfnamefont {H.}~\bibnamefont {Kamano}},\
  and\ \bibinfo {author} {\bibfnamefont {T.}~\bibnamefont {Sato}},\ }\bibfield
  {title} {\bibinfo {title} {{Dynamical coupled-channels model for
  neutrino-induced meson productions in resonance region}},\ }\href
  {https://doi.org/10.1103/PhysRevD.92.074024} {\bibfield  {journal} {\bibinfo
  {journal} {Phys. Rev. D}\ }\textbf {\bibinfo {volume} {92}},\ \bibinfo
  {pages} {074024} (\bibinfo {year} {2015})},\ \Eprint
  {https://arxiv.org/abs/1506.03403} {arXiv:1506.03403 [hep-ph]} \BibitemShut
  {NoStop}%
\bibitem [{\citenamefont {Thorpe}\ \emph {et~al.}(2021)\citenamefont {Thorpe},
  \citenamefont {Nowak}, \citenamefont {Niewczas}, \citenamefont {Sobczyk},\
  and\ \citenamefont {Juszczak}}]{Thorpe:2020tym}%
  \BibitemOpen
  \bibfield  {author} {\bibinfo {author} {\bibfnamefont {C.}~\bibnamefont
  {Thorpe}}, \bibinfo {author} {\bibfnamefont {J.}~\bibnamefont {Nowak}},
  \bibinfo {author} {\bibfnamefont {K.}~\bibnamefont {Niewczas}}, \bibinfo
  {author} {\bibfnamefont {J.~T.}\ \bibnamefont {Sobczyk}},\ and\ \bibinfo
  {author} {\bibfnamefont {C.}~\bibnamefont {Juszczak}},\ }\bibfield  {title}
  {\bibinfo {title} {{Second class currents, axial mass, and nuclear effects in
  hyperon production}},\ }\href {https://doi.org/10.1103/PhysRevC.104.035502}
  {\bibfield  {journal} {\bibinfo  {journal} {Phys. Rev. C}\ }\textbf {\bibinfo
  {volume} {104}},\ \bibinfo {pages} {035502} (\bibinfo {year} {2021})},\
  \Eprint {https://arxiv.org/abs/2010.12361} {arXiv:2010.12361 [hep-ph]}
  \BibitemShut {NoStop}%
\bibitem [{\citenamefont {Sobczyk}\ \emph {et~al.}(2019)\citenamefont
  {Sobczyk}, \citenamefont {Rocco}, \citenamefont {Lovato},\ and\ \citenamefont
  {Nieves}}]{Sobczyk:2019uej}%
  \BibitemOpen
  \bibfield  {author} {\bibinfo {author} {\bibfnamefont {J.~E.}\ \bibnamefont
  {Sobczyk}}, \bibinfo {author} {\bibfnamefont {N.}~\bibnamefont {Rocco}},
  \bibinfo {author} {\bibfnamefont {A.}~\bibnamefont {Lovato}},\ and\ \bibinfo
  {author} {\bibfnamefont {J.}~\bibnamefont {Nieves}},\ }\bibfield  {title}
  {\bibinfo {title} {{Weak Production of Strange and Charmed Ground-State
  Baryons in Nuclei}},\ }\href {https://doi.org/10.1103/PhysRevC.99.065503}
  {\bibfield  {journal} {\bibinfo  {journal} {Phys. Rev. C}\ }\textbf {\bibinfo
  {volume} {99}},\ \bibinfo {pages} {065503} (\bibinfo {year} {2019})},\
  \Eprint {https://arxiv.org/abs/1901.10192} {arXiv:1901.10192 [nucl-th]}
  \BibitemShut {NoStop}%
\bibitem [{\citenamefont {Lalakulich}\ \emph {et~al.}(2012)\citenamefont
  {Lalakulich}, \citenamefont {Gallmeister},\ and\ \citenamefont
  {Mosel}}]{Lalakulich:2012gm}%
  \BibitemOpen
  \bibfield  {author} {\bibinfo {author} {\bibfnamefont {O.}~\bibnamefont
  {Lalakulich}}, \bibinfo {author} {\bibfnamefont {K.}~\bibnamefont
  {Gallmeister}},\ and\ \bibinfo {author} {\bibfnamefont {U.}~\bibnamefont
  {Mosel}},\ }\bibfield  {title} {\bibinfo {title} {{Neutrino- and
  antineutrino-induced reactions with nuclei between 1 and 50 GeV}},\ }\href
  {https://doi.org/10.1103/PhysRevC.86.014607} {\bibfield  {journal} {\bibinfo
  {journal} {Phys. Rev. C}\ }\textbf {\bibinfo {volume} {86}},\ \bibinfo
  {pages} {014607} (\bibinfo {year} {2012})},\ \Eprint
  {https://arxiv.org/abs/1205.1061} {arXiv:1205.1061 [nucl-th]} \BibitemShut
  {NoStop}%
\bibitem [{\citenamefont {De~Vries}\ \emph {et~al.}(1987)\citenamefont
  {De~Vries}, \citenamefont {De~Jager},\ and\ \citenamefont
  {De~Vries}}]{DeVries:1987atn}%
  \BibitemOpen
  \bibfield  {author} {\bibinfo {author} {\bibfnamefont {H.}~\bibnamefont
  {De~Vries}}, \bibinfo {author} {\bibfnamefont {C.~W.}\ \bibnamefont
  {De~Jager}},\ and\ \bibinfo {author} {\bibfnamefont {C.}~\bibnamefont
  {De~Vries}},\ }\bibfield  {title} {\bibinfo {title} {{Nuclear charge and
  magnetization density distribution parameters from elastic electron
  scattering}},\ }\href {https://doi.org/10.1016/0092-640X(87)90013-1}
  {\bibfield  {journal} {\bibinfo  {journal} {Atom. Data Nucl. Data Tabl.}\
  }\textbf {\bibinfo {volume} {36}},\ \bibinfo {pages} {495} (\bibinfo {year}
  {1987})}\BibitemShut {NoStop}%
\bibitem [{\citenamefont {Vidana}\ \emph {et~al.}(1998)\citenamefont {Vidana},
  \citenamefont {Polls}, \citenamefont {Ramos},\ and\ \citenamefont
  {Hjorth-Jensen}}]{Vidana:1998ed}%
  \BibitemOpen
  \bibfield  {author} {\bibinfo {author} {\bibfnamefont {I.}~\bibnamefont
  {Vidana}}, \bibinfo {author} {\bibfnamefont {A.}~\bibnamefont {Polls}},
  \bibinfo {author} {\bibfnamefont {A.}~\bibnamefont {Ramos}},\ and\ \bibinfo
  {author} {\bibfnamefont {M.}~\bibnamefont {Hjorth-Jensen}},\ }\bibfield
  {title} {\bibinfo {title} {{Hyperon properties in finite nuclei using
  realistic Y N interactions}},\ }\href
  {https://doi.org/10.1016/S0375-9474(98)00599-5} {\bibfield  {journal}
  {\bibinfo  {journal} {Nucl. Phys. A}\ }\textbf {\bibinfo {volume} {644}},\
  \bibinfo {pages} {201} (\bibinfo {year} {1998})},\ \Eprint
  {https://arxiv.org/abs/nucl-th/9805032} {arXiv:nucl-th/9805032} \BibitemShut
  {NoStop}%
\bibitem [{\citenamefont {Rodr\'\i{}guez-S\'anchez}\ \emph
  {et~al.}(2018)\citenamefont {Rodr\'\i{}guez-S\'anchez}, \citenamefont
  {David}, \citenamefont {Hirtz}, \citenamefont {Cugnon},\ and\ \citenamefont
  {Leray}}]{Rodriguez-Sanchez:2018oqv}%
  \BibitemOpen
  \bibfield  {author} {\bibinfo {author} {\bibfnamefont {J.~L.}\ \bibnamefont
  {Rodr\'\i{}guez-S\'anchez}}, \bibinfo {author} {\bibfnamefont {J.~C.}\
  \bibnamefont {David}}, \bibinfo {author} {\bibfnamefont {J.}~\bibnamefont
  {Hirtz}}, \bibinfo {author} {\bibfnamefont {J.}~\bibnamefont {Cugnon}},\ and\
  \bibinfo {author} {\bibfnamefont {S.}~\bibnamefont {Leray}},\ }\bibfield
  {title} {\bibinfo {title} {{Constraining the ${\Lambda}$-nucleus potential
  within the Li\`ege intranuclear cascade model}},\ }\href
  {https://doi.org/10.1103/PhysRevC.98.021602} {\bibfield  {journal} {\bibinfo
  {journal} {Phys. Rev. C}\ }\textbf {\bibinfo {volume} {98}},\ \bibinfo
  {pages} {021602} (\bibinfo {year} {2018})}\BibitemShut {NoStop}%
\bibitem [{\citenamefont {Hirtz}\ \emph {et~al.}(2020)\citenamefont {Hirtz},
  \citenamefont {David}, \citenamefont {Boudard}, \citenamefont {Cugnon},
  \citenamefont {Leray}, \citenamefont {Leya}, \citenamefont
  {Rodr\'\i{}guez-S\'anchez},\ and\ \citenamefont {Schnabel}}]{Hirtz:2019rqe}%
  \BibitemOpen
  \bibfield  {author} {\bibinfo {author} {\bibfnamefont {J.}~\bibnamefont
  {Hirtz}}, \bibinfo {author} {\bibfnamefont {J.~C.}\ \bibnamefont {David}},
  \bibinfo {author} {\bibfnamefont {A.}~\bibnamefont {Boudard}}, \bibinfo
  {author} {\bibfnamefont {J.}~\bibnamefont {Cugnon}}, \bibinfo {author}
  {\bibfnamefont {S.}~\bibnamefont {Leray}}, \bibinfo {author} {\bibfnamefont
  {I.}~\bibnamefont {Leya}}, \bibinfo {author} {\bibfnamefont {J.~L.}\
  \bibnamefont {Rodr\'\i{}guez-S\'anchez}},\ and\ \bibinfo {author}
  {\bibfnamefont {G.}~\bibnamefont {Schnabel}},\ }\bibfield  {title} {\bibinfo
  {title} {{Strangeness production in the new version of the Li\`ege
  intranuclear cascade model}},\ }\href
  {https://doi.org/10.1103/PhysRevC.101.014608} {\bibfield  {journal} {\bibinfo
   {journal} {Phys. Rev. C}\ }\textbf {\bibinfo {volume} {101}},\ \bibinfo
  {pages} {014608} (\bibinfo {year} {2020})},\ \Eprint
  {https://arxiv.org/abs/1909.02246} {arXiv:1909.02246 [nucl-th]} \BibitemShut
  {NoStop}%
\bibitem [{\citenamefont {Harada}\ and\ \citenamefont
  {Hirabayashi}(2023)}]{Harada:2023otu}%
  \BibitemOpen
  \bibfield  {author} {\bibinfo {author} {\bibfnamefont {T.}~\bibnamefont
  {Harada}}\ and\ \bibinfo {author} {\bibfnamefont {Y.}~\bibnamefont
  {Hirabayashi}},\ }\bibfield  {title} {\bibinfo {title} {{Production spectra
  with a \ensuremath{\Sigma}\ensuremath{-} hyperon in
  (\ensuremath{\pi}\ensuremath{-},K+) reactions on light to heavy nuclei}},\
  }\href {https://doi.org/10.1103/PhysRevC.107.054611} {\bibfield  {journal}
  {\bibinfo  {journal} {Phys. Rev. C}\ }\textbf {\bibinfo {volume} {107}},\
  \bibinfo {pages} {054611} (\bibinfo {year} {2023})}\BibitemShut {NoStop}%
\bibitem [{\citenamefont {Oset}\ and\ \citenamefont
  {Vicente-Vacas}(1986)}]{Oset:1986qd}%
  \BibitemOpen
  \bibfield  {author} {\bibinfo {author} {\bibfnamefont {E.}~\bibnamefont
  {Oset}}\ and\ \bibinfo {author} {\bibfnamefont {M.~J.}\ \bibnamefont
  {Vicente-Vacas}},\ }\bibfield  {title} {\bibinfo {title} {{Inclusive ($\pi$,
  2 $\pi$) Reactions in Nuclei}},\ }\href
  {https://doi.org/10.1016/0375-9474(86)90110-7} {\bibfield  {journal}
  {\bibinfo  {journal} {Nucl. Phys. A}\ }\textbf {\bibinfo {volume} {454}},\
  \bibinfo {pages} {637} (\bibinfo {year} {1986})}\BibitemShut {NoStop}%
\bibitem [{\citenamefont {Singh}\ \emph {et~al.}(2006)\citenamefont {Singh},
  \citenamefont {Sajjad~Athar},\ and\ \citenamefont {Ahmad}}]{Singh:2006br}%
  \BibitemOpen
  \bibfield  {author} {\bibinfo {author} {\bibfnamefont {S.~K.}\ \bibnamefont
  {Singh}}, \bibinfo {author} {\bibfnamefont {M.}~\bibnamefont
  {Sajjad~Athar}},\ and\ \bibinfo {author} {\bibfnamefont {S.}~\bibnamefont
  {Ahmad}},\ }\bibfield  {title} {\bibinfo {title} {{Nuclear effects in
  neutrino induced coherent pion production at K2K and MiniBooNE}},\ }\href
  {https://doi.org/10.1103/PhysRevLett.96.241801} {\bibfield  {journal}
  {\bibinfo  {journal} {Phys. Rev. Lett.}\ }\textbf {\bibinfo {volume} {96}},\
  \bibinfo {pages} {241801} (\bibinfo {year} {2006})},\ \Eprint
  {https://arxiv.org/abs/nucl-th/0601045} {arXiv:nucl-th/0601045} \BibitemShut
  {NoStop}%
\bibitem [{\citenamefont {Zhang}\ and\ \citenamefont
  {Serot}(2012)}]{Zhang:2012xi}%
  \BibitemOpen
  \bibfield  {author} {\bibinfo {author} {\bibfnamefont {X.}~\bibnamefont
  {Zhang}}\ and\ \bibinfo {author} {\bibfnamefont {B.~D.}\ \bibnamefont
  {Serot}},\ }\bibfield  {title} {\bibinfo {title} {{Coherent
  Neutrinoproduction of Photons and Pions in a Chiral Effective Field Theory
  for Nuclei}},\ }\href {https://doi.org/10.1103/PhysRevC.86.035504} {\bibfield
   {journal} {\bibinfo  {journal} {Phys. Rev. C}\ }\textbf {\bibinfo {volume}
  {86}},\ \bibinfo {pages} {035504} (\bibinfo {year} {2012})},\ \Eprint
  {https://arxiv.org/abs/1208.1553} {arXiv:1208.1553 [nucl-th]} \BibitemShut
  {NoStop}%
\bibitem [{\citenamefont {Oset}\ and\ \citenamefont
  {Salcedo}(1987)}]{Oset:1987re}%
  \BibitemOpen
  \bibfield  {author} {\bibinfo {author} {\bibfnamefont {E.}~\bibnamefont
  {Oset}}\ and\ \bibinfo {author} {\bibfnamefont {L.~L.}\ \bibnamefont
  {Salcedo}},\ }\bibfield  {title} {\bibinfo {title} {{$\Delta$ Selfenergy in
  Nuclear Matter}},\ }\href {https://doi.org/10.1016/0375-9474(87)90185-0}
  {\bibfield  {journal} {\bibinfo  {journal} {Nucl. Phys. A}\ }\textbf
  {\bibinfo {volume} {468}},\ \bibinfo {pages} {631} (\bibinfo {year}
  {1987})}\BibitemShut {NoStop}%
\bibitem [{\citenamefont {Salcedo}\ \emph {et~al.}(1988)\citenamefont
  {Salcedo}, \citenamefont {Oset}, \citenamefont {Vicente-Vacas},\ and\
  \citenamefont {Garcia-Recio}}]{Salcedo:1987md}%
  \BibitemOpen
  \bibfield  {author} {\bibinfo {author} {\bibfnamefont {L.~L.}\ \bibnamefont
  {Salcedo}}, \bibinfo {author} {\bibfnamefont {E.}~\bibnamefont {Oset}},
  \bibinfo {author} {\bibfnamefont {M.~J.}\ \bibnamefont {Vicente-Vacas}},\
  and\ \bibinfo {author} {\bibfnamefont {C.}~\bibnamefont {Garcia-Recio}},\
  }\bibfield  {title} {\bibinfo {title} {{Computer Simulation of Inclusive Pion
  Nuclear Reactions}},\ }\href {https://doi.org/10.1016/0375-9474(88)90310-7}
  {\bibfield  {journal} {\bibinfo  {journal} {Nucl. Phys. A}\ }\textbf
  {\bibinfo {volume} {484}},\ \bibinfo {pages} {557} (\bibinfo {year}
  {1988})}\BibitemShut {NoStop}%
\bibitem [{\citenamefont {Andreopoulos}\ \emph {et~al.}(2010)\citenamefont
  {Andreopoulos} \emph {et~al.}}]{Andreopoulos:2009rq}%
  \BibitemOpen
  \bibfield  {author} {\bibinfo {author} {\bibfnamefont {C.}~\bibnamefont
  {Andreopoulos}} \emph {et~al.},\ }\bibfield  {title} {\bibinfo {title} {{The
  GENIE Neutrino Monte Carlo Generator}},\ }\href
  {https://doi.org/10.1016/j.nima.2009.12.009} {\bibfield  {journal} {\bibinfo
  {journal} {Nucl. Instrum. Meth. A}\ }\textbf {\bibinfo {volume} {614}},\
  \bibinfo {pages} {87} (\bibinfo {year} {2010})},\ \Eprint
  {https://arxiv.org/abs/0905.2517} {arXiv:0905.2517 [hep-ph]} \BibitemShut
  {NoStop}%
\bibitem [{\citenamefont {Hewes}\ \emph {et~al.}(2021)\citenamefont {Hewes}
  \emph {et~al.}}]{DUNE:2021tad}%
  \BibitemOpen
  \bibfield  {author} {\bibinfo {author} {\bibfnamefont {V.}~\bibnamefont
  {Hewes}} \emph {et~al.} (\bibinfo {collaboration} {DUNE}),\ }\bibfield
  {title} {\bibinfo {title} {{Deep Underground Neutrino Experiment (DUNE) Near
  Detector Conceptual Design Report}},\ }\href
  {https://doi.org/10.3390/instruments5040031} {\bibfield  {journal} {\bibinfo
  {journal} {Instruments}\ }\textbf {\bibinfo {volume} {5}},\ \bibinfo {pages}
  {31} (\bibinfo {year} {2021})},\ \Eprint {https://arxiv.org/abs/2103.13910}
  {arXiv:2103.13910 [physics.ins-det]} \BibitemShut {NoStop}%
\bibitem [{\citenamefont {Abed~Abud}\ \emph {et~al.}(2022)\citenamefont
  {Abed~Abud} \emph {et~al.}}]{DUNE:2022yni}%
  \BibitemOpen
  \bibfield  {author} {\bibinfo {author} {\bibfnamefont {A.}~\bibnamefont
  {Abed~Abud}} \emph {et~al.} (\bibinfo {collaboration} {DUNE}),\ }\bibfield
  {title} {\bibinfo {title} {{A Gaseous Argon-Based Near Detector to Enhance
  the Physics Capabilities of DUNE}},\ }\href@noop {} {\  (\bibinfo {year}
  {2022})},\ \Eprint {https://arxiv.org/abs/2203.06281} {arXiv:2203.06281
  [hep-ex]} \BibitemShut {NoStop}%
\end{thebibliography}%

\end{document}